\documentclass[showpacs,preprintnumbers,superscriptaddress,amsmath,amssymb,floatfix,nofootinbib,aps]{revtex4}

\usepackage[dvips]{graphicx,graphics}
\usepackage{dcolumn}% Align table columns on decimal point
\usepackage{bm}
\usepackage{epsfig}

\newcommand{\MeV}{\;\text{MeV}}

\begin{document}

\title{Number of the QCD critical points with neutral color superconductivity}
\author{Zhao Zhang}\email{zhaozhang@pku.org.cn}
\affiliation{Yukawa Institute for Theoretical Physics, Kyoto
University, Kyoto 606-8502, Japan}
\affiliation{Department of Physics, Kyoto University, Kyoto 606-8502, Japan}
\author{Kenji Fukushima}\email{fuku@yukawa.kyoto-u.ac.jp}
\affiliation{Yukawa Institute for Theoretical Physics, Kyoto
University, Kyoto 606-8502, Japan}
\author{Teiji Kunihiro}\email{kunihiro@ruby.scphys.kyoto-u.ac.jp}
\affiliation{Department of Physics, Kyoto University, Kyoto 606-8502, Japan}

\pacs{12.38.Aw, 11.10.Wx, 11.30.Rd, 12.38.Gc}

\begin{abstract}

 We investigate the effect of the electric-charge neutrality in
 $\beta$-equilibrium on the chiral phase transition by solving the
 chiral and diquark condensates in the two-flavor Nambu--Jona-Lasinio
 model.  We demonstrate that the electric-charge neutrality plays a
 similar role as the repulsive vector interaction; they both weaken
 the first-order chiral phase transition in the high-density and
 low-temperature region.  The first-order chiral phase transition is
 not affected, however, at finite temperatures where the diquark
 condensate melts.  In this way the chiral phase transition could be
 second-order at intermediate temperatures if the diquark effects
 overwhelm the chiral dynamics, while the first-order transition may
 survive at lower and higher temperatures.  The number of the critical
 points appearing on the phase diagram can vary from zero to three,
 which depends on the relative strength of the chiral and diquark
 couplings.  We systematically study the possibility of the phase
 structure with multiple QCD critical points and evaluate the Meissner
 screening  mass to confirm that our conclusion is not overturned by
 chromomagnetic instability.

\end{abstract}
\preprint{YITP-08-69, KUNS-2154}
\maketitle

%%%%%%%%%%   INTRODUCTION   %%%%%%%%%%

\section{INTRODUCTION}

  It is generally believed that quantum chromodynamics (QCD) exhibits
a rich phase structure in extreme environment such as high temperature
and high baryon density.  In the last decade the color-superconducting
(CSC) phase has attracted lots of theoretical interests and triggered
extensive studies of dense and cold quark
matter~\cite{WilczekReview,RischkeReview,BuballaReview,AlfordReview}.
At asymptotically high density that justifies the perturbative QCD
calculations the color-flavor locked (CFL) phase~\cite{CFL} has been
established as the ground state of quark matter.  There, the pairing
energy, the critical temperature, the screening properties, the
collective excitation energy, and so on are well understood from the
first-principle calculations.

  In a practical sense, however, the accessible baryon density in
nature would be, at most, ten times the normal nuclear density even in
the interior of the compact stellar objects.  In terms of the quark
chemical potential the corresponding value should be less than
$500\MeV$ in reality and thus other MeV energy scales such as the
current and dynamical quark masses, the electric chemical potential,
etc take part into the dynamics.  We are yet far from thorough
understanding on the phase structure in this density region which is
commonly referred to as the intermediate density.  Going down from the
CFL phase, we have to rely on chiral effective theories to unveil the
possible CSC phases and to draw the boundary lines on the phase
diagram which separate different CSC
states~\cite{BuballaReview,AlfordReview}.

  Recently, an interesting proposal has been made for the plausible
phase structure in the intermediate density
region~\cite{Hatsuda:2006ps}.  That is, there may exist a new QCD
critical point (i.e.\ the terminal of the first-order phase
boundary) induced by the $\mathrm{U_A}(1)$-breaking vertex.  This
speculation is based on a general Ginzburg-Landau theory in terms
of the order parameter fields constrained by QCD symmetries.
There, the three-flavor anomaly term generates the coupling
between $\langle\bar{\psi}\psi\rangle$ (chiral condensate),
$\langle\psi\psi\rangle$ (diquark condensate), and
$\langle\bar{\psi}\bar{\psi}\rangle$ (antidiquark condensate).  It
has been then argued that the resultant crossover of chiral
restoration at small temperature embodies the hadron-quark
continuity hypothesis~\cite{Continuity}.  Whether the new critical
point appears with reasonable parameter set or not needs further
investigation. Maybe, to settle the situation without ambiguity,
we should wait for future developments in lattice QCD simulations
at finite density.

  In a different context, before discussed in
Ref.~\cite{Hatsuda:2006ps}, the appearance of another critical
point at low temperature was pointed out in
Ref.~\cite{KitazawaVector} within the two-flavor
Nambu--Jona-Lasinio (NJL) model.  The crucial ingredient in
Ref.~\cite{KitazawaVector} is the four-fermion interaction in the
vector channel as well as in the scalar one.  In this case, the vector
interaction diminishes the first-order chiral phase
transition~\cite{Asakawa,Klimt,Buballa1996}, which allows for
enhanced competition between the chiral and diquark condensates in
the widened coexisting phase.  For some choices of the NJL model
parameter two critical points show up along the phase boundary
that signifies crossover, first-order transition, and crossover again
with increasing temperature.  We note that a similar phase diagram
with two critical points is suggested in the lattice calculation of
two-color QCD~\cite{Lattice2color}, though it seems to have not
been established yet.  As mentioned in Ref.~\cite{KitazawaVector},
the vector-channel interaction could give a possible explanation
for this two-color phase structure with two critical points.

  In this paper we will reveal another mechanism leading to multiple
QCD critical points; the electric-charge neutrality realized in
$\beta$ equilibrium can make chiral restoration being smooth
crossover at low temperature, so that an analogous situation to
Refs.~\cite{Hatsuda:2006ps,KitazawaVector} takes place.  It is
worth remarking here that imposing the electric neutrality has a
similar effect on the thermodynamic potential to introducing the
vector-channel interaction.  One can understand this in terms of
the electric chemical potential, $\mu_e$, which we define such
that $\mu_e>0$  not for positron but electron.  In the model
treatment without gauge fields a nonzero $\mu_e$ mimics the role
of $A_0$ to neutralize the system.  The point is that $\mu_e$
enters the dynamics like the vector-channel interaction;  a
vector-channel interaction
$-G_{\text{V}}(\bar{\psi}\gamma^\mu\psi)^2$ induces an effective
(renormalized) chemical potential~\cite{Asakawa} which takes a
form,
\begin{equation}
 \mu_{\text{R}} = \mu-2G_{\text{V}}\cdot\rho_q \,.
\label{vector}
\end{equation}
Here $G_\text{V}>0$ represents the repulsive vector coupling constant
and $\rho_q$ is the quark number density,
$\langle\bar{\psi}\gamma^0\psi\rangle$.  For two-flavor QCD $\mu_e$
leads to a mismatch between the $u$-quark chemical potential
$\mu_u=\mu-\frac{2}{3}\mu_e$ and the $d$-quark one
$\mu_d=\mu+\frac{1}{3}\mu_e$.  We shall use a notation, $\bar{\mu}$,
to denote the average chemical potential for $u$ and $d$ quarks
resulting in
\begin{equation}
 \bar{\mu} = \mu-\frac{1}{6}\mu_e \,.
\label{eq:average_mu}
\end{equation}
Comparing the forms of $\mu_{\text{R}}$ and $\bar{\mu}$ above, one
may well anticipate that $\mu_e\neq0$ can be taken in effect as a
repulsive vector coupling for the bulk properties, in addition to
keeping the electric neutrality.  In physics terms, the electric
chemical potential realizing a finite electric-charge density in
part plays a role as a baryon chemical potential on its own.  This
situation is drastically different from the massless three-flavor
case where the electric charge generator happens to be traceless
so that no coupling between the electric-charge and baryon
fluctuations arises (i.e. no term proportional to $\mu_e$ arises
in $\bar{\mu}$ in this case).

  In the best of our knowledge the direct coupling between the quark
density and the electric chemical potential has drawn only little
attention so far, though the vector-channel interaction and the
electric neutrality have been investigated in separate contexts.  In
view of the results in Ref.~\cite{KitazawaVector}, which is produced
solely by the vector interaction, it is natural to expect that $\mu_e$
may also have a significant impact on both the chiral and CSC phase
transitions.

  The purpose of this paper is to investigate this issue seriously and
depict an intuitive picture which opens a possibility to drive more
QCD critical points than only one.  We shall demonstrate our idea in
the framework of the two-flavor NJL model for concreteness.  Indeed,
the competition between the chiral and diquark dynamics results in
zero, one, two, and three critical points depending on the relative
strength of the chiral and diquark couplings.

  The paper is organized as follows.  In Sec.~\ref{sec:model}, the
model is introduced and the formalism is presented.  The numerical
results and discussions are given in Sec.~\ref{sec:results}.  The
final section is devoted to the summary and concluding remarks.

%%%%%%%%%%   MODEL AND FORMALISM   %%%%%%%%%%

\section{MODEL AND FORMALISM}
\label{sec:model}

  We will explain the choice of the effective model, the model
parameters, and the resulting thermodynamic potential in order.  Since
we adopt a standard description by the NJL model, the experts could
skip to our results in Sec.~\ref{sec:results}.

%%%   Model   %%%

\subsection{Model}

  Varieties of NJL-type models have been extensively used to
investigate the CSC phase transition at moderate and large
density~\cite{BuballaReview} as well as at zero
density~\cite{Klevansky:1992,Hatsuda:1994}.  For the two-flavor case,
the commonly used Lagrangian of the NJL model reads
\begin{equation}
 \mathcal{L}_{\text{NJL}} = \bar{\psi}\left(i\gamma^{\mu}\partial_{\mu}
  -\widehat{m}_0\right)\psi + \mathcal{L}_{\bar{q}q}+\mathcal{L}_{qq} \,,
\label{lagrangian}
\end{equation}
where the chiral interaction part is
\begin{equation}
 \mathcal{L}_{\bar{q}q} = G\left[(\bar{\psi}\psi\right)^2
  +\left(\bar{\psi}i\gamma_5 \vec{\tau}\psi)^2\right] \,,
\label{scalar}
\end{equation}
and the diquark part which is relevant to the mean-field condensate
(i.e.\ the spin, flavor, and color are all antisymmetric) is
\begin{equation}
 \mathcal{L}_{qq} = H\left[(\bar{\psi}C\gamma_5\tau_2\lambda_A
  \bar{\psi}^T)(\psi^T C\gamma_5\tau_2\lambda_A\psi) \right] \,.
\end{equation}
Here, $C=i\gamma_0\gamma_2$ stands for the Dirac charge
conjugation matrix, and $G$ and $H$ are the coupling constants for
the mesonic and diquark channels.  The current quark mass matrix
is given by $\widehat{m}=\text{diag}(m_u,m_d)$ in two flavors and
we shall work in the isospin symmetric limit with $m_u=m_d=m$.  We
note that $\lambda_A$'s are the antisymmetric Gell-Mann matrices
(i.e.\ $A$ runs over $ 2, 5, 7$ only) for the color SU(3) group
and $\vec{\tau}$'s are the Pauli matrices in flavor space.

  For simplicity, the coupling constant in the vector channel is set
to zero in this paper.  The effect of the electric-charge neutrality
on the chiral phase transition with including the Polyakov loop
dynamics and the nonzero vector interaction will be reported in our
future work~\cite{future}.

  It deserves noting here that the scalar four-fermion interaction
$\mathcal{L}_{\bar{q}q}$ in general consists of two types of different
interactions~\cite{Klevansky:1992,Hatsuda:1994,Hatsuda:1985ey}, that is,
\begin{align}
 \mathcal{L}_1 &= G_1 \left[(\bar{\psi}\psi)^2
  +(\bar{\psi}\vec{\tau}\psi)^2+(\bar{\psi}i\gamma_5\psi)^2
  +(\bar{\psi}i\gamma_5\vec{\tau}\psi)^2 \right], \\
 \mathcal{L}_2 &= G_2 \left[(\bar{\psi}\psi)^2
  -(\bar{\psi}\vec{\tau}\psi)^2-(\bar{\psi}i\gamma_5\psi)^2
  +(\bar{\psi}i\gamma_5\vec{\tau}\psi)^2 \right].
\end{align}
Both interaction terms have the symmetry of
$\mathrm{SU(2)_L}\times\mathrm{SU(2)_R}\times\mathrm{U(1)}$, while
the axial symmetry, $\mathrm{U_A(1)}$, remains only in
$\mathcal{L}_1$. The $\mathrm{U_A(1)}$ breaking part
$\mathcal{L}_2$ belongs to the instanton-induced (two-flavor
't~Hooft) interaction.  In the mean-field level the constituent
quark masses in the presence of $\mathcal{L}_1$ and
$\mathcal{L}_2$ are
\begin{equation}
 M_u = m - 4G_1\langle{\bar{u}u}\rangle
   - 4G_2\langle{\bar{d}d}\rangle \,, \qquad
 M_d = m - 4G_1\langle{\bar{d}d}\rangle
   - 4G_2\langle{\bar{u}u}\rangle \,.
\label{Massgap}
\end{equation}
Therefore, in general, $M_u\neq M_d$ if there exists a chemical
potential mismatch between $u$ quarks and $d$ quarks by $\mu_e$
and thus $\langle\bar{u}u\rangle\neq\langle\bar{d}d\rangle$.  If
we introduce a parameter $\alpha$ to relate $G_1$ and $G_2$ to $G$
in a way that
\begin{equation}
 G_1 = (1-\alpha)G, \qquad  G_2 = \alpha{G},
\label{eq:alpha}
\end{equation}
we notice that the standard Lagrangian~(\ref{scalar}) corresponds
to the case of $\alpha=0.5$.  In such a case, as is clear from
Eq.~(\ref{Massgap}), the constituent mass of $u$ quarks is always
identical to that of $d$ quarks regardless of a difference in
$\mu_u$ and $\mu_d$.  Once we get ready to proceed to the
numerical calculations, in Sec.~\ref{sec:alpha}, we will check the
dependence on $\alpha$ in a simple case without diquark
condensation.  In any case, because this paper aims to illustrate
a general feature in the phase structure, we shall stick to the
simplest choice $\alpha=0.5$ which makes no difference in the
qualitative picture.

  Then, there are four model parameters left; the current quark mass
$m$ of $u$ and $d$ quarks, the coupling constants $G$ and $H$, and
the three-momentum cutoff $\Lambda$.  In this work, we take the
same parameters as in Ref.~\cite{Zhang:2007} which are fixed so as
to reproduce the three physical quantities in vacuum;  the pion
mass $m_\pi\approx 140\MeV$, the pion decay constant $f_\pi\approx
94\MeV$, and the chiral condensate
$\langle\bar{u}u\rangle=\langle\bar{d}d\rangle\approx-(251\MeV)^3$
with
\begin{equation}
 m=5.5\,\mathrm{MeV},\qquad G=5.04\,\mathrm{GeV}^{-2},\qquad
 \Lambda=0.651\,\mathrm{GeV} \,.
\end{equation}
The corresponding constituent quark mass in vacuum is $325.5\MeV$
for this set of the model parameter.  The standard value of the
ratio $H/G$ is $3/4=0.75$, which is deduced by the Fierz
transformation from the local current-current interaction.  In
this paper, we rather treat this ratio as a free parameter and
shall perform a systematic survey.

%%%   Thermodynamic Potential with Neutrality Condition   %%%

\subsection{Thermodynamic Potential with Neutrality Condition}

  In general, the quark chemical potential matrix $\hat{\mu}$ takes
the form~\cite{Alford:2002kj}
\begin{equation}
 \hat{\mu} = \mu - \mu_e Q + \mu_3T_3 + \mu_8T_8,
\end{equation}
where $\mu$ is the quark chemical potential (i.e.\ one third of
the baryon chemical potential), $\mu_e$ is the chemical potential
associated with the (negative) electric charge, and $\mu_3$ and
$\mu_8$ represent the color chemical potentials corresponding to
the Cartan subalgebra in color SU(3) space.  The explicit form of
the electric-charge matrix is
$Q=\text{diag}(\frac{2}{3},-\frac{1}{3})$ in flavor space, and the
color charge matrices are
$T_3=\text{diag}(\frac{1}{2},-\frac{1}{2},0)$ and
$T_8=\text{diag}(\frac{1}{3},\frac{1}{3}, -\frac{2}{3})$ in color
space.  The chemical potentials for different quarks are listed
below;
\begin{equation}
 \begin{split}
&\mu_{ru} = \mu-\tfrac{2}{3}\mu_e+\tfrac{1}{2}\mu_3+\tfrac{1}{3}\mu_8 \,,
 \qquad
 \mu_{gu} = \mu-\tfrac{2}{3}\mu_e-\tfrac{1}{2}\mu_3+\tfrac{1}{3}\mu_8 \,,\\
&\mu_{rd} = \mu+\tfrac{1}{3}\mu_e+\tfrac{1}{2}\mu_3+\tfrac{1}{3}\mu_8 \,,
 \qquad
 \mu_{gd} = \mu+\tfrac{1}{3}\mu_e-\tfrac{1}{2}\mu_3+\tfrac{1}{3}\mu_8 \,,\\
&\mu_{bu} = \mu-\tfrac{2}{3}\mu_e-\tfrac{2}{3}\mu_8 \,, \qquad\qquad\quad
 \mu_{bd} = \mu+\tfrac{1}{3}\mu_e-\tfrac{2}{3}\mu_8 \,.
 \end{split}
\end{equation}
The four-quark interactions develop a dynamical quark mass with
nonzero chiral condensate as
\begin{equation}
 M = m-\sigma = m - 2G\langle{\bar{\psi}\psi\rangle} \,,
\end{equation}
while the diquark condensate $\Delta$ and antidiquark condensate
$\Delta^*$ could appear at high enough baryon density.  Here, we
follow the common treatment for two-flavor CSC that the blue
quarks do not take part in the Cooper pairing.

  Using the standard bosonization technique, the mean-field
thermodynamic potential in the NJL model with the diquark degrees
of freedom as well as the electron contribution takes the
following form:
\begin{equation}
 \Omega = \frac{{\sigma^2}}{4G}+\frac{{\Delta^2}}{4H}
  -\frac{1}{12\pi^2}\left(\mu_e^4+2\pi^2T^2\mu_e^2
            +\frac{7\pi^4}{15}T^4\right)
  -T\sum_n\int\frac{d^3p}{\left(2\pi\right)^3}\mathrm{Tr}\ln
  \frac{{S}_{\text{MF}}^{-1}\left(i\omega_n,\vec{p}\,\right)}{T} \,,
\label{omega}
\end{equation}
where the sum runs over the Matsubara frequency
$\omega_n=(2n+1)\pi{T}$ and Tr is taken over color, flavor, and Dirac
indices.  The inverse quark propagator matrix including both the
chiral and diquark condensates in the Nambu-Gor'kov formalism is then
given by
\begin{equation}
 S^{-1}_{\mathrm{MF}}(i\omega_n,\vec{p}) = \bigg(\begin{array}{cc}
  [{G_0^{+}}]^{-1} & \Delta\gamma_5\tau_2\lambda_2 \\
  -\Delta^*\gamma_5\tau_2\lambda_2 &
  [{G_0^{-}}]^{-1} \end{array}\bigg) \,,
\end{equation}
with
\begin{equation}
 [{G_0^{\pm}}]^{-1}=\gamma_0(i\omega_n\pm\hat{\mu})
  -\vec{\gamma}\cdot\vec{p}-\widehat{m} \,.
\end{equation}
Taking the Matsubara sum, we can express the thermodynamic potential
as usual as
\begin{equation}
 \begin{split}
 &\Omega(\mu_e,\mu_3,\mu_8,\sigma,\Delta;\mu,T) \\
 &=\frac{\sigma^2}{4G}+\frac{\Delta^2}{4H}
  -\frac{1}{12\pi^2}\left(\mu_e^4+2\pi^2T^2\mu_e^2
   +\frac{7\pi^4}{15}T^4\right)
  -\sum_{i=1}^{12}\int\frac{d^3p}{(2\pi)^3}\{E_i+2T\ln(1+e^{-E_i/T})\},
 \end{split}
\label{eqn:therp}
\end{equation}
with the dispersion relations for six quasiparticles [that is, 2
flavors $\times$ 3 colors;  the spin degeneracy is already taken
into account in Eq.~(\ref{eqn:therp})] and $6$
quasianti-particles. The unpaired blue quarks have the following
four energy dispersion relations,
\begin{equation}
 E_{bu} = E - \mu_{bu} \,\quad, \bar{E}_{bu} = E + \mu_{bu} \,\quad
 E_{bd} = E - \mu_{bd} \,\quad, \bar{E}_{bd} = E + \mu_{bd}
\end{equation}
with $E=\sqrt{\vec{p}^2+M^2}$.  In the $rd$-$gu$ quark sector with
pairing we can find the four dispersion relations,
\begin{equation}
 \begin{split}
 E_{\text{$rd$-$gu$}}^{\pm} = E_\Delta \pm \tfrac{1}{2}(\mu_{rd}-\mu_{gu})
  = E_\Delta \pm \tfrac{1}{2}(\mu_e+\mu_3) \,,\\
 \bar{E}_{\text{$rd$-$gu$}}^{\pm} = \bar{E}_\Delta \pm
  \tfrac{1}{2}(\mu_{rd}-\mu_{gu})
  = \bar{E}_\Delta \pm \tfrac{1}{2}(\mu_e+\mu_3) \,,
 \end{split}
\end{equation}
and the $ru$-$gd$ sector has another four as
\begin{equation}
 \begin{split}
 E_{\text{$ru$-$gd$}}^{\pm} = E_\Delta \pm \tfrac{1}{2}(\mu_{ru}-\mu_{gd})
  = E_\Delta \mp \tfrac{1}{2}(\mu_e-\mu_3) \,,\\
 \bar{E}_{\text{$ru$-$gd$}}^{\pm} = \bar{E}_\Delta \pm
  \tfrac{1}{2}(\mu_{ru}-\mu_{gd})
  = \bar{E}_\Delta \mp \tfrac{1}{2}(\mu_e-\mu_3) \,,
 \end{split}
\end{equation}
where $E_\Delta=\sqrt{(E-\bar{\mu})^2+\Delta^2}$ and
$\bar{E}_\Delta=\sqrt{(E+\bar{\mu})^2+\Delta^2}$.  The average
chemical potential is defined by
\begin{equation}
 \bar{\mu} = \frac{\mu_{rd}+\mu_{gu}}{2}
 = \frac{\mu_{ru}+\mu_{gd}}{2} = \mu-\frac{\mu_e}{6}
 + \frac{\mu_8}{3} \,.
\label{Average}
\end{equation}

  For the two-flavor CSC case, the color charge corresponding to the
matrix $T_3$ is always zero since the color SU(2) symmetry is left
unbroken for red and green quarks.  That means $\mu_3=0$.  In contrast
to that in the NJL model, nontrivial coupling to the Polyakov loop
might induce a nonzero $\mu_3$ in the case of the Polyakov loop
augmented NJL (PNJL) model~\cite{Roessner:2006xn,PlApp8}, which is
beyond the current scope.

  Since we know that $\mu_8$ is much smaller than
$\mu_e$~\cite{Alford:2002kj,Huang} to neutralize two-flavor CSC
matter, the positive $\mu_e/6$ is overwhelming in Eq.~(\ref{Average})
so that we can neglect $\mu_8$ in the numerical calculation.  The
average chemical potential then amounts to Eq.~(\ref{eq:average_mu}).

  Minimizing the thermodynamic potential~(\ref{eqn:therp}), we can
solve the mean fields $\sigma$ and $\Delta$ together with the chemical
potential $\mu_e$ from
\begin{equation}
 \frac{\partial\Omega}{\partial\sigma}=
 \frac{\partial\Omega}{\partial\Delta}=
 \frac{\partial\Omega}{\partial\mu_e}=0 \,.
\end{equation}

%%%   Dependence on $\alpha$ and the Constituent Quark Mass Difference   %%%

\subsection{Dependence on $\alpha$ and the Constituent Quark Mass Difference}
\label{sec:alpha}

\begin{figure}
 \includegraphics[width=\textwidth]{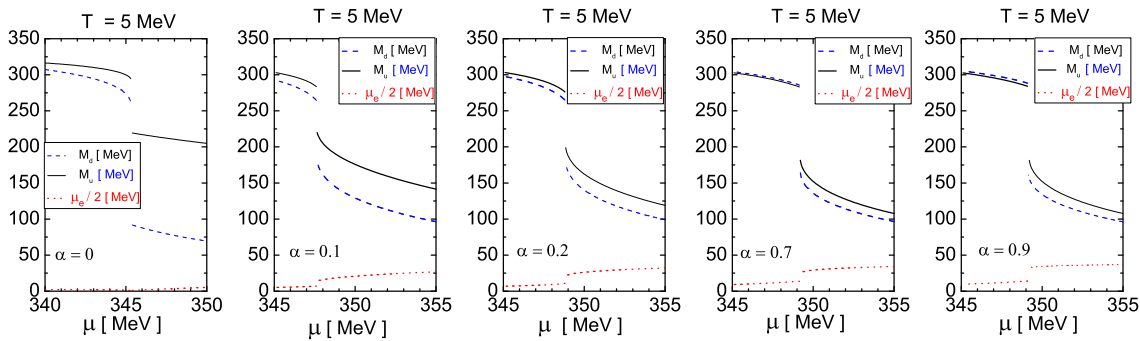}
 \caption{The constituent quark mass difference between $u$ and $d$
   quarks for $\alpha=0$, $0.1$, $0.2$, $0.7$, $0.9$, where $\alpha$
   is a parameter to indicate the flavor-mixing interaction defined in
   Eq.~(\ref{eq:alpha}).  All results are obtained under the electric
   charge neutrality and without diquark condensation.}
 \label{fig:massmismatch}
\end{figure}

\begin{figure}
 \includegraphics[width=0.6\textwidth]{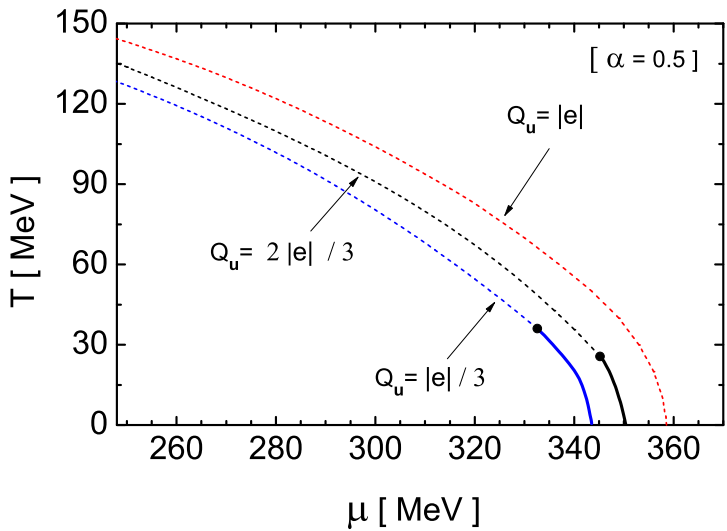}
 \caption{The phase diagram for the chiral phase transition under the
   electric-charge neutrality.  The solid line (dashed line)
   represents the first-order transition (smooth crossover) and the
   filled circle dot locates the QCD critical point.}
 \label{fig:effectue}
\end{figure}

  For the cases with $\alpha\neq 0.5$, as we have mentioned, $M_u$
should be different from $M_d$ in the presence of nonzero $\mu_e$,
which is apparent in Eq.~(\ref{Massgap}).
Figure~\ref{fig:massmismatch} shows the dependence on $\alpha$ in
the behavior of $M_u$, $M_d$, and $\mu_e$ as a function of $\mu$
at a fixed temperature $T=5\MeV$.  We do not take account of the
diquark condensate for the moment.  From the figure we note the
following two points:  First, $\mu_e$ becomes larger with
increasing $\alpha$, and the first-order phase transition tends to
occur at a higher chemical potential.  This means that the value
of $\mu_e$, which plays a role similar to the vector interaction
in the chiral phase transition, is sensitive to the magnitude of
the flavor-mixing interaction.  Second, in contrast to their
absolute values of masses, we see that the mass differences
between $u$ and $d$ quarks are sizable for $\alpha=0$ and
$\alpha=0.1$, while the mass difference becomes minor for
$\alpha\geq 0.2$ .

  It should be noted that the strength of the flavor-mixing
interaction which originates from the $\mathrm{U_A}(1)$ anomaly may be
small in comparison with the $\mathrm{U}(2)\times\mathrm{U}(2)$
symmetric part, and it would be intriguing to explore its effect on
$M_u$ and $M_d$.  We stress, however, that we should perform such
studies for the three-flavor case where the quantitative effect of the
$\mathrm{U_A}(1)$ anomaly is more clearly seen in the $\eta$-$\eta'$
system (for instance~\cite{foot:1}).  In the present two-flavor
analysis, therefore, we shall only consider $\alpha=0.5$ for
elucidating the effect of $\mu_e$ on the chiral and CSC phase
transitions.  Actually, Fig.~\ref{fig:massmismatch} shows that we can
reasonably ignore the difference between $M_u$ and $M_d$ induced by
$\mu_e$ unless $\alpha<0.2$.

  It is known that as the coupling constant $G_\text{V}$ of the vector
interaction increases, the whole critical or crossover line of the
chiral phase transition shifts toward larger chemical potential
and the critical point moves toward smaller temperature and larger
chemical potential, which disappears eventually at a large value
of $G_\text{V}$~\cite{Asakawa,Klimt,Buballa1996,KitazawaVector}.
Let us show that the $\mu_e$ affects the phase diagram in the
similar way as $G_\text{V}$, which is anticipated from
Eqs.~(\ref{vector}) and (\ref{eq:average_mu}).  To demonstrate it
clearly, we try to enhance the effect of $\mu_e$ artificially by
varying the $u$-quark electric charge $Q_u$ by hand as $Q_u=|e|$,
$2|e|/3$, $|e|/3$, while keeping the $d$ quarks unchanged:  The
case of $Q_u=|e|/3$ corresponds to the situation with no net
effect on the average chemical potential.  The real world is
$Q_u=2|e|/3$.  We can induce a further large net shift in the
average chemical potential by taking $Q_u=|e|$.
Figure~\ref{fig:effectue} shows that the critical point shifts
from $(T,\mu)=(36\MeV,\,332\MeV)$ for $Q_u=|e|/3$ to
$(T,\mu)=(24\MeV,\,345\MeV)$ for $Q_u=2|e|/3$, and eventually
disappears from the phase diagram when we choose $Q_u=|e|$.  These
results are quite reminiscent of the effects of changing $G_V$
discussed in the
literature~\cite{Asakawa,Klimt,Buballa1996,KitazawaVector}.

%%%%%%%%%%   NUMERICAL RESULTS AND DISCUSSIONS   %%%%%%%%%%

\section{NUMERICAL RESULTS AND DISCUSSIONS}
\label{sec:results}

  In this section, we shall discuss the effect of electric-charge
neutrality in $\beta$-equilibrium on the phase structure with the CSC
phase as well as with the chiral transition taken into consideration.
We shall show that the $\mu_e$ induced by the neutrality constraint
gives rise to a phase structure with multiple critical points, and the
number of the critical points can be zero, one, two, and three.

\begin{figure}
 \includegraphics[width=10cm]{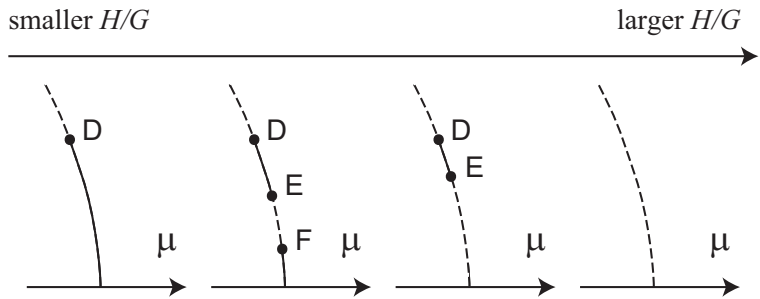}
 \caption{The schematic change of the phase structure with increasing
   $H/G$ from the left to the right.  The solid and dashed lines
   represent the first-order transition and crossover, respectively.
   The number of the critical points depends on $H/G$.}
 \label{fig:schematic}
\end{figure}

  Before presenting our numerical results, we shall give an intuitive
account of the mechanism which causes the phase diagram with multiple
critical points.  We depict a schematic sketch in
Fig.~\ref{fig:schematic} which is useful to explain what would be
anticipated in advance.  There is a terminal point of the first-order
phase boundary, that is the QCD critical point located at
$\textsf{D}$, as long as the coupling ratio $H/G$ is small and hence
the neutrality constraint is not significant.  When the coupling ratio
$H/G$ becomes substantially large, the critical line of the
first-order phase transition is breached by the induced $\mu_e$ in the
CSC phase, and there appear two more edges in the critical line of the
first-order transition;  the new critical points are denoted as
$\textsf{E}$ and $\textsf{F}$.  Here we notice that there exist three
critical points as a whole.  So to speak, the first-order line
attached to $\textsf{F}$ is a ``survivor'' transition that surpasses
the diquark effect at sufficiently low temperature where the
first-order chiral transition remains strong.  The first-order line
$\textsf{D}$-$\textsf{E}$ is, on the other hand, to be regarded as a
``remnant'' where the diquark condensate almost melts and would hardly
affect the chiral transition.  For a larger $H/G$, the survivor may be
gone and then the two critical points $\textsf{D}$ and $\textsf{E}$
are left on the phase diagram.  If we further increase $H/G$, all the
first-order transitions of chiral restoration and all the critical
points are washed away eventually.

  In the subsequent subsections we shall present numerical results and
see what is described above is actually the case.  For convenience
we shall adopt the same notations as those in
Ref.~\cite{Hatsuda:2006ps} to distinguish the different regions on
the $T$-$\mu$ phase diagram;  NG, CSC, COE, and NOR refer to the
hadronic (Nambu-Goldstone) phase with $\sigma\neq0$ and
$\Delta=0$, the color-superconducting phase with $\Delta\neq0$ and
$\sigma=0$, the coexisting phase with $\sigma\neq0$ and $\Delta\neq0$,
and the normal phase with $\sigma=\Delta=0$, respectively, though they
have exact meaning only in the chiral limit.  In fact, as seen
from our results such as Figs.~\ref{fig:gapmedium}~(b),
\ref{fig:gapr80}~(b), \ref{fig:gapr875}, and so on, $M$ stays
$10\sim100\MeV$ even in CSC but near COE.

%%%   The Case of Intermediate Diquark Coupling   %%%

\subsection{The Case of Intermediate Diquark Coupling}

\begin{figure}
\hspace{-.05\textwidth}
\begin{minipage}[t]{.4\textwidth}
\includegraphics*[width=\textwidth]{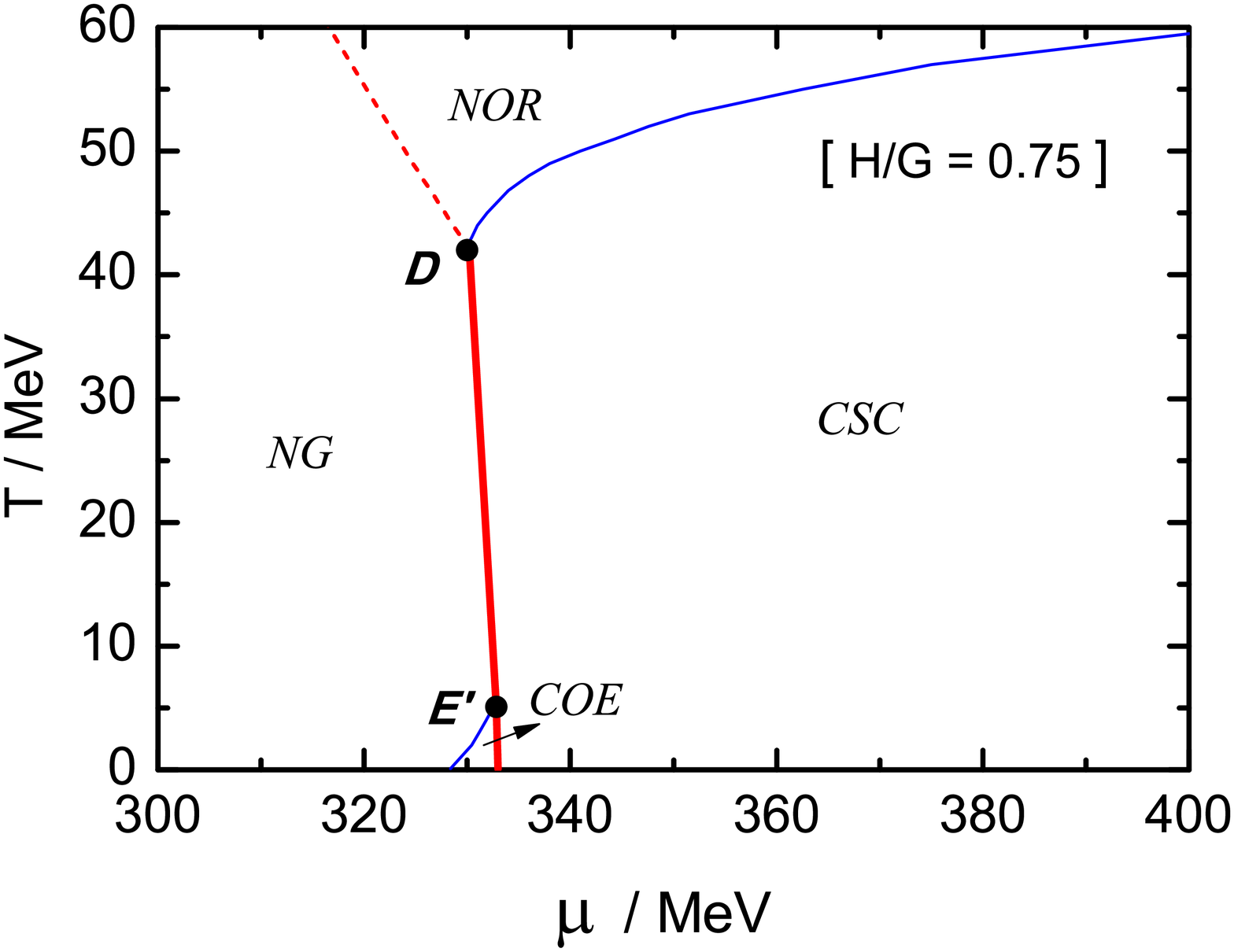}
\centerline{(a)}
\end{minipage}
\hspace{.05\textwidth}
\begin{minipage}[t]{.4\textwidth}
\includegraphics*[width=\textwidth]{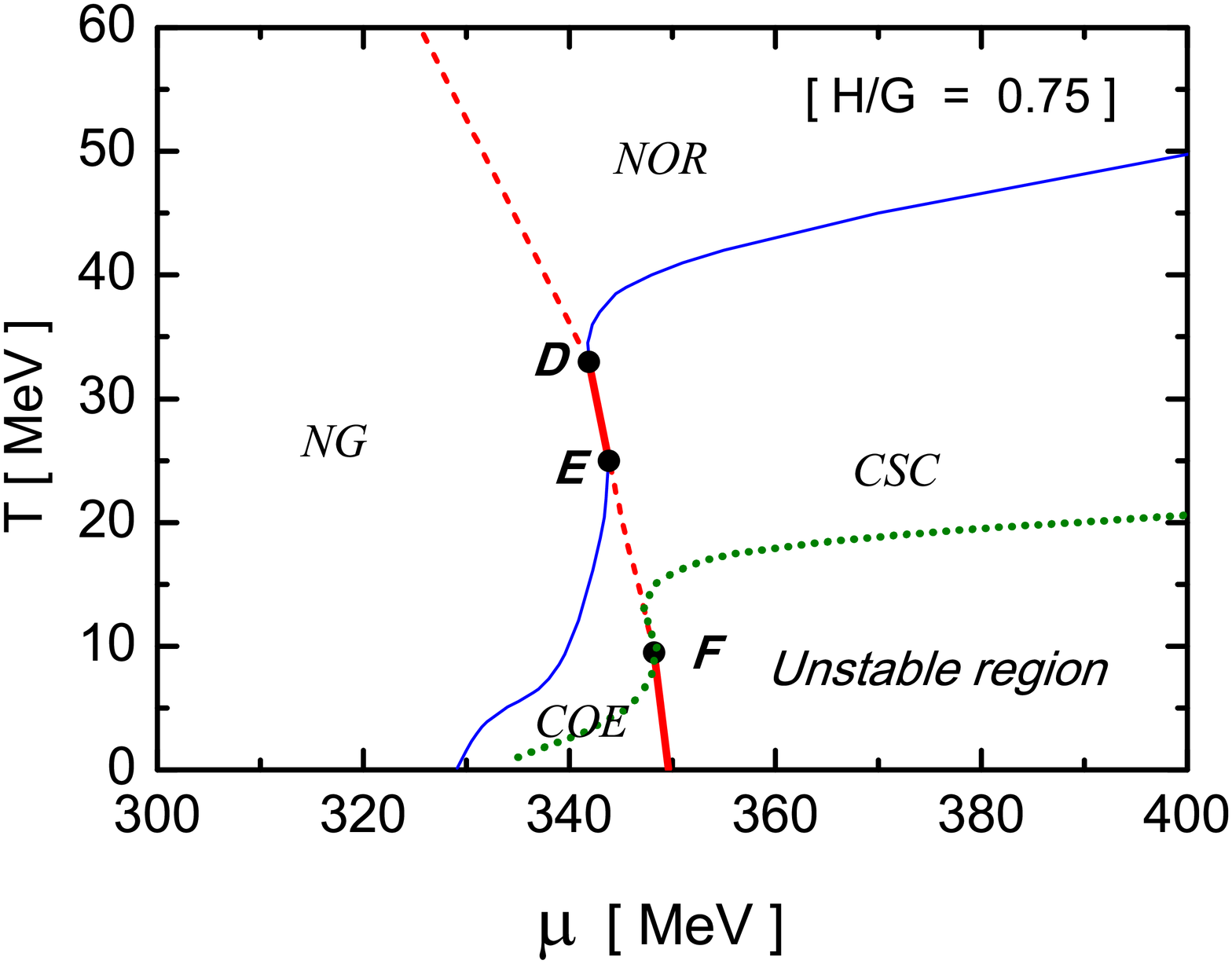}
\centerline{(b)}
\end{minipage}
\caption{The phase diagrams for the intermediate coupling constant
  $H/G=0.75$ including the diquark condensate.  The left (a) (right
  (b)) panel corresponds to the case without (with) enforcing the
  charge neutrality.  In the left figure the unstable region is
  indicated by the dotted curve (see the discussion in
  Sec.~\ref{sec:Meissner}).}
\label{fig:medium}
\end{figure}

\begin{figure}
\hspace{.05\textwidth}
\begin{minipage}[t]{\textwidth}
\includegraphics[width=\textwidth]{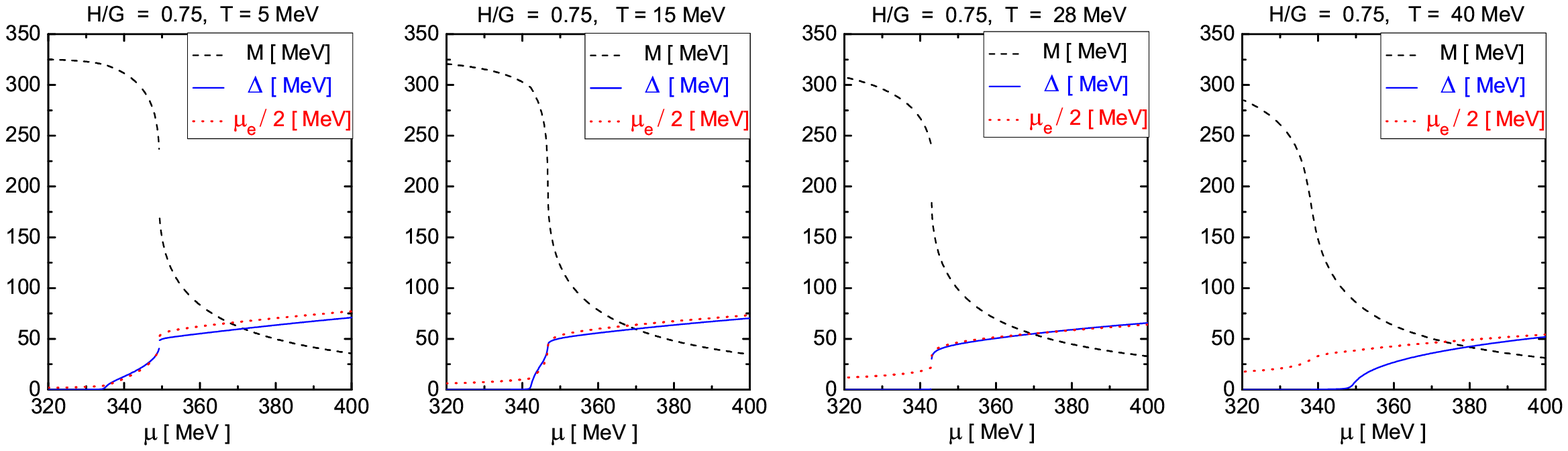}
\centerline{(a)}
\vspace*{1mm}
\end{minipage}
\hspace{-.05\textwidth}
\begin{minipage}[t]{\textwidth}
\includegraphics[width=\textwidth]{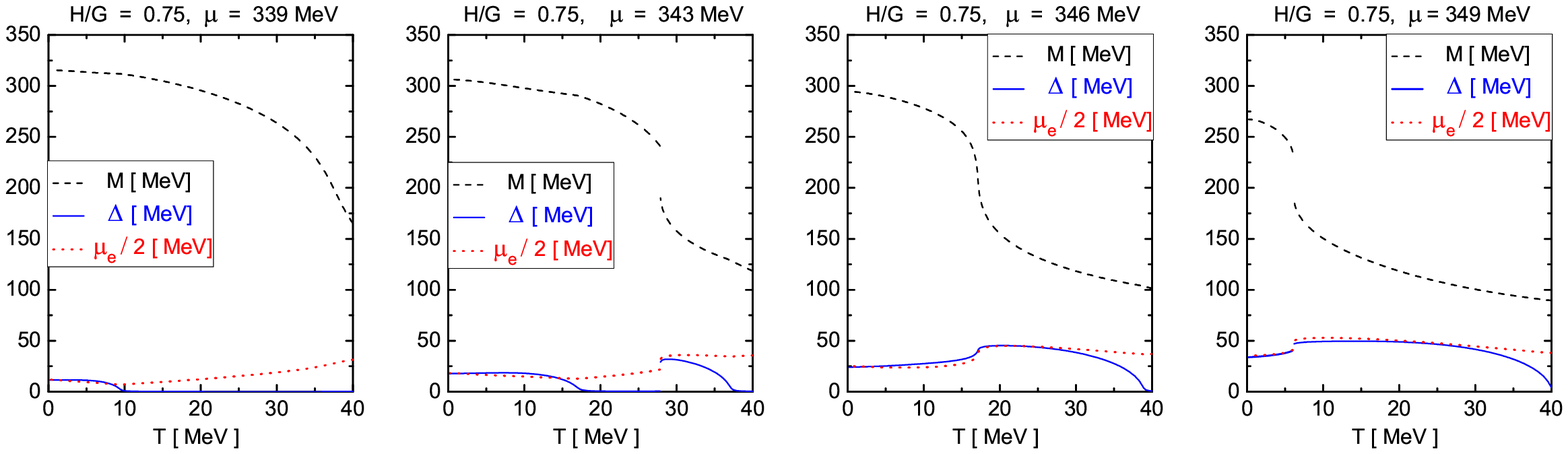}
\centerline{(b)}
\end{minipage}
\caption{$M$, $\Delta$, and $\mu_e$ as functions of the chemical
  potential in (a) and the temperature in (b) for $H/G=0.75$ under the
  condition of electric-charge neutrality.}
\label{fig:gapmedium}
\end{figure}

  We first consider the standard ratio $H/G=0.75$, which is usually
referred to as the ``intermediate'' diquark coupling strength.  Two
phase diagrams with and without the charge-neutrality constraint are
presented in Fig.~\ref{fig:medium};  henceforth, the thick solid
curve, the dashed curve, and the thin solid curve in the $T$-$\mu$
plane stand for the critical lines of first-order phase transition,
smooth crossover, and second-order phase transition, respectively.
Since the magnitude of $\mu_8$ is very small~\cite{Huang}, as we have
mentioned, we ignore it in the numerical calculation.  We have checked
that nonzero $\mu_8$ has only a minor effect on our main results.

  Both panels in Fig.~\ref{fig:medium} show that all the four phases,
i.e., NG, CSC, COE, and NOR are realized in  the phase diagram.  The
upper triple point at which the NG, CSC, and NOR phases encounter
happens to be the critical point $\textsf{D}$ for the chiral-to-CSC
phase transition.

  Figure~\ref{fig:medium}~(a) shows that the chiral phase transition
keeps of first order at low temperatures even with the emergence of a
COE phase when the charge-neutrality condition is not imposed.  We
remark that similar phase diagrams were obtained in the previous
works~\cite{KitazawaVector,Rapp,Carter,RadomMatrix} without imposing
the charge-neutrality constraint in the NJL model, the instanton-based
models, and the random matrix model.

  In contrast to Fig.~\ref{fig:medium}~(a), one finds an
unconventional phase structure in Fig.~\ref{fig:medium}~(b) where the
charge-neutrality constraint is imposed:  The would-be critical line
for the first-order transition is terminated at two points
$\textsf{E}$ and $\textsf{F}$, between which the phase transition
becomes crossover from COE to CSC.\ \ Hence there exist three distinct
critical points, \textsf{D}, \textsf{E}, and \textsf{F}, as a whole.
Although a possible phase structure with {\em two} critical points
 was suggested previously~\cite{KitazawaVector,Hatsuda:2006ps},
the present work is the first model study that shows possible
existence of {\em three} critical points in the QCD phase diagram.

  In addition to the appearance of the three-critical-point structure,
the following points are notable in Fig.~\ref{fig:medium}~(b):
\vspace{0.3em}\\
\noindent i)~ The critical line for the first-order chiral phase
transition in the low-temperature region is shifted considerably
toward higher quark chemical potential.  Furthermore, the critical
point \textsf{D} is somewhat shifted to lower $T$ and higher
$\mu$; $(T,\mu)=(42\MeV,\,330\MeV)$ $\rightarrow$
$(33\MeV,\,341\MeV)$.  These are to be understood as the same
effect of $\mu_e$ as we mentioned in the explanation of
Fig.~\ref{fig:effectue}.
\vspace{0.3em}\\
\noindent
ii)~
The coexisting region is widened both in the $T$ and $\mu$ directions.
The lower triple point of NG, COE, and CSC, labeled as \textsf{E}' in
Fig.~\ref{fig:medium}~(a) and \textsf{E} in (b), moves toward higher
$T$ and $\mu$;
$(T,\mu)=(5\MeV,\,333\MeV)$ $\rightarrow$
$(25\MeV,\,343\MeV)$.
\vspace{0.3em}

  It is not difficult to understand the emergence of a smooth
crossover for the chiral phase transition in the low-temperature
region of the phase diagram when the diquark pairing is taken into
account.  The positive $\mu_e$ first considerably reduces the
discontinuity in the constituent quark mass at the first-order
transition, and then the competition between the chiral and
diquark condensates leads to the complete disappearance of the
discontinuity. This situation is similar to what happens with the
vector interaction as described in Ref.~\cite{KitazawaVector};  in
Ref.~\cite{KitazawaVector}, though the charge-neutrality condition
is not imposed, a similar phase diagram to ours is obtained and
the enlargement of the coexisting phase is attributed to the
enhanced competition between the chiral and diquark condensations
by the vector interaction.

  Now let us discuss the mechanism for realizing the
three-critical-point structure shown in Fig.~\ref{fig:medium}.  For
this sake, we also show, in Fig.~\ref{fig:gapmedium}, the $\mu$ and
$T$ dependence of $M$, $\Delta$, and $\mu_e$ in the upper and lower
panels, respectively.  We see the nature of the chiral phase
transition from low to high temperatures in order.
\vspace{0.3em}\\
\noindent 1)~ In the low-temperature region below the critical
point \textsf{F}, we have a first-order transition from the COE to
CSC phase, as shown in Fig.~\ref{fig:medium}~(b).  This feature is
also clearly exhibited in the first panel of
Fig.~\ref{fig:gapmedium}~(a) and the fourth panel in
Fig.~\ref{fig:gapmedium}~(b) both of which show discontinuous
jumps in the physical quantities.  This result can be interpreted
as follows; in this low-temperature region, the chiral condensate
even in the COE phase has a rather large value and dominates over
the diquark condensate at present diquark coupling $H/G=0.75$.  We
should notice that the diquark condensate smears the Fermi surface
like at finite temperature, which in turn tends to make the chiral
transition weak~\cite{KitazawaVector}.  In short, the effect of
the diquark condensate is not yet strong enough in this
temperature region to convert the first-order chiral phase
transition to a smooth crossover.  We identify this remaining
first-order transition a survivor of the chiral transition which
should be first order without the diquark condensate.
\vspace{0.3em}\\
\noindent
2)~ Figure~\ref{fig:medium}(b) shows that at intermediate temperatures
between \textsf{E} and \textsf{F}, the chiral transition becomes
smooth.  The underlying mechanism for this may be understood from an
unusual temperature dependence of the diquark condensate in the
relevant chemical potential region.  In fact, as shown in the third
panel of Fig.~\ref{fig:gapmedium}~(b), the diquark condensate
increases as $T$ is raised for a fixed $\mu$.  This is due to the
combined effects of the charge-neutrality constraint and the
temperature~\cite{Huang}:  The neutrality constraint causes mismatched
Fermi spheres for the quarks involved in the pairing, which unfavors
the pairing, especially at small temperatures.  At larger temperature,
the mismatched Fermi surfaces are smeared enough to allow for the
significant number of the quarks involved in the pairing, and
hence the diquark condensate can develop.  Besides, the chiral
condensate decreases in a monotonic way with increasing temperature,
and a smaller quark mass favors the formation of the diquark
condensate because of a larger Fermi surface at a fixed $\mu$.  Thus
the effect of the diquark condensate may overwhelm the chiral
condensate at intermediate temperatures, which suppresses
discontinuity in the chiral phase transition which turns to
crossover.
\vspace{0.3em}\\
\noindent 3)~ In the still higher temperature region, we find that
the phase transition comes back to first order from NG to CSC,
which is between \textsf{D} and \textsf{E} in
Fig.~\ref{fig:medium}~(b).  These features can be understood as
follows.  First, we note that \textsf{D} is attached to the
critical line where $\Delta$ melts, and the diquark condensate
decreases along \textsf{D}-\textsf{E} with the increasing
temperature, as shown in the second panel of
Fig.~\ref{fig:medium}~(b).  It means that the chiral condensate
dominates over the diquark pairing in this region.  Thus the
original feature of the chiral transition without the diquark
condensation is intact there, and COE does not appear.  We see
that the chiral restoration remains of first order as shown in the
third panel of Fig.~\ref{fig:gapmedium}~(a) and the second panel
of Fig.~\ref{fig:gapmedium}~(b).  In short, the first-order
transition in this region is a remnant of the chiral restoration
existing without the effect of the diquark pairing.
\vspace{0.3em}\\
\noindent
4)~ For even higher temperatures than \textsf{D}, the chiral
transition is a crossover and the condensates only show smooth
behavior as seen in the fourth panel of Fig.~\ref{fig:gapmedium}~(a)
and the first panel in Fig.~\ref{fig:gapmedium}~(b) where the
second-order nature of the CSC transition is also exhibited.

%%%   The Case of Strong Diquark Coupling   %%%

\subsection{The Case of Strong Diquark Coupling}

\begin{figure}
\hspace{-.05\textwidth}
\begin{minipage}[t]{0.4\textwidth}
\includegraphics[width=\textwidth]{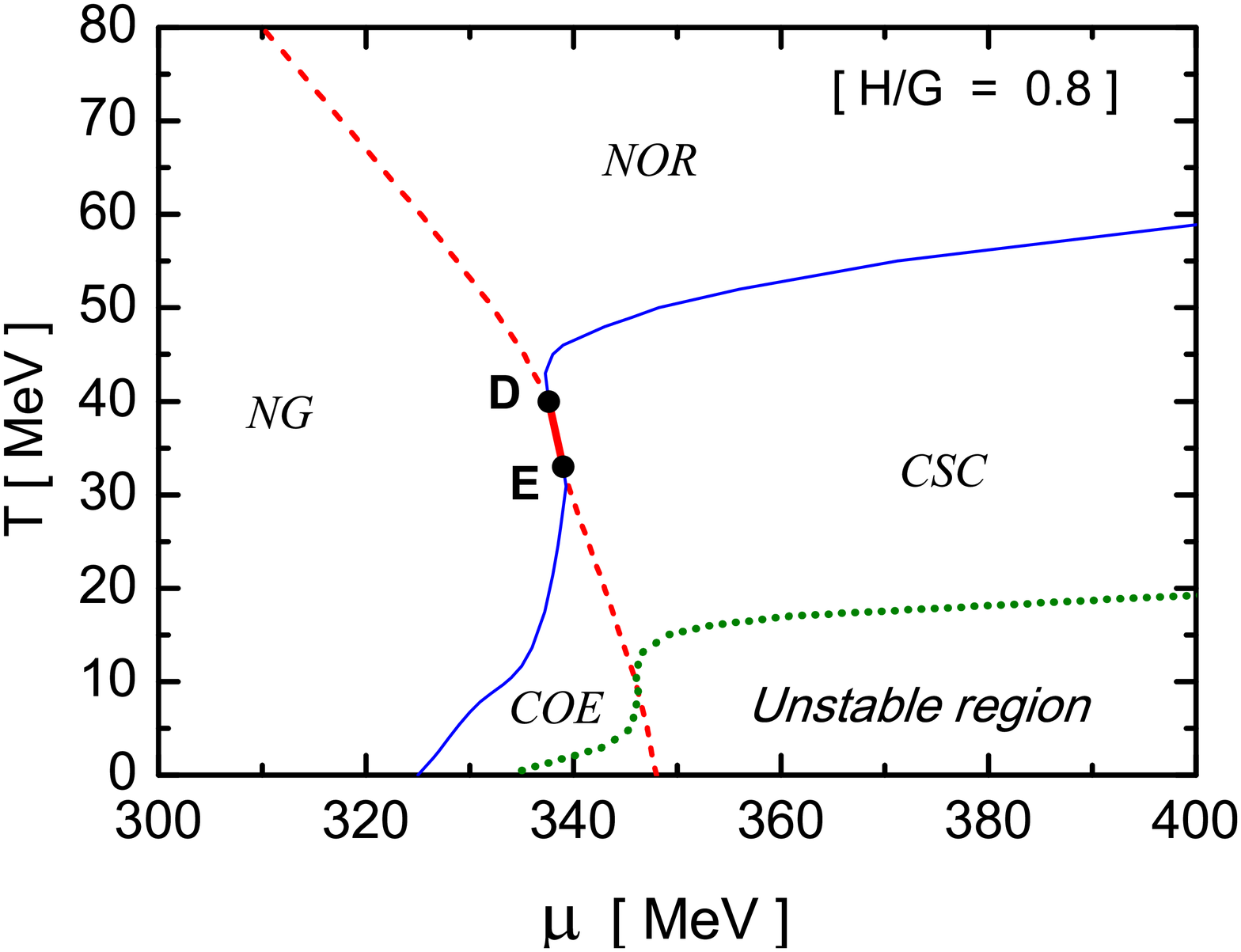}
\centerline{(a)}
\end{minipage}
\hspace{.05\textwidth}
\begin{minipage}[t]{0.4\textwidth}
\includegraphics[width=\textwidth]{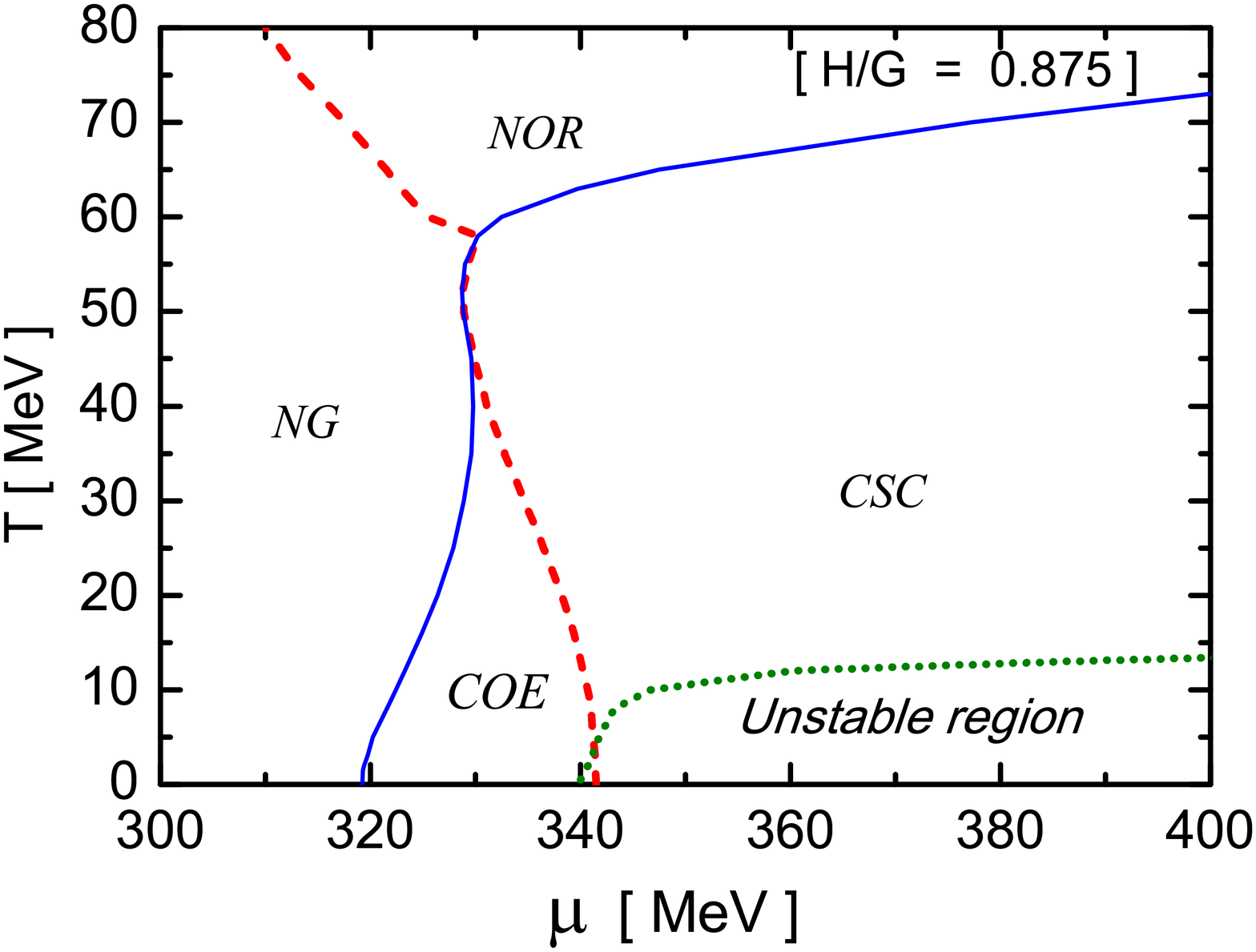}
\centerline{(b)}
\end{minipage}
\caption{The phase diagrams for the strong diquark couplings
$H/G=0.8$ (a) and $H/G=0.875$ (b) under electric-charge neutrality.
  The unstable region is indicated by the dotted curve as previously.}
\label{fig:phasestrong}
\end{figure}

\begin{figure}
\hspace{-.01\textwidth}
\begin{minipage}[t]{0.75\textwidth}
\includegraphics[width=\textwidth]{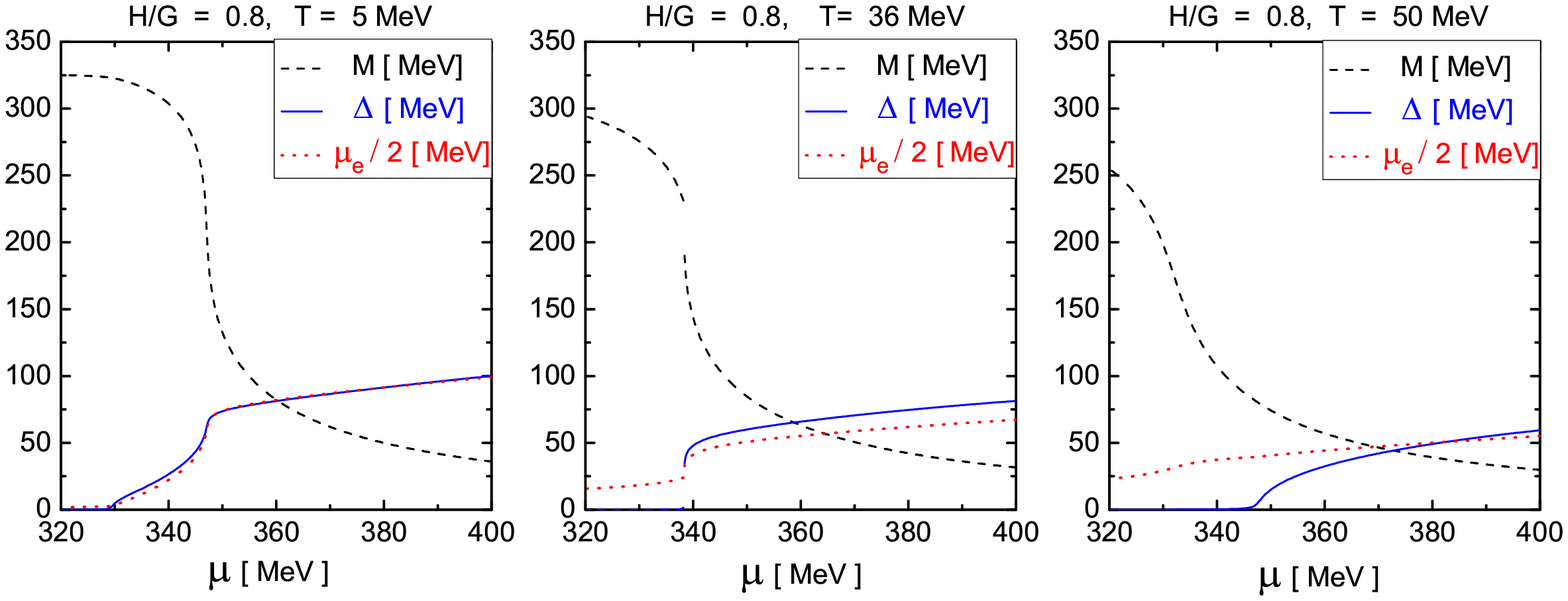}
\centerline{(a)}
\vspace*{1mm}
\end{minipage}
\hspace{-.05\textwidth}
\begin{minipage}[t]{\textwidth}
\includegraphics[width=\textwidth]{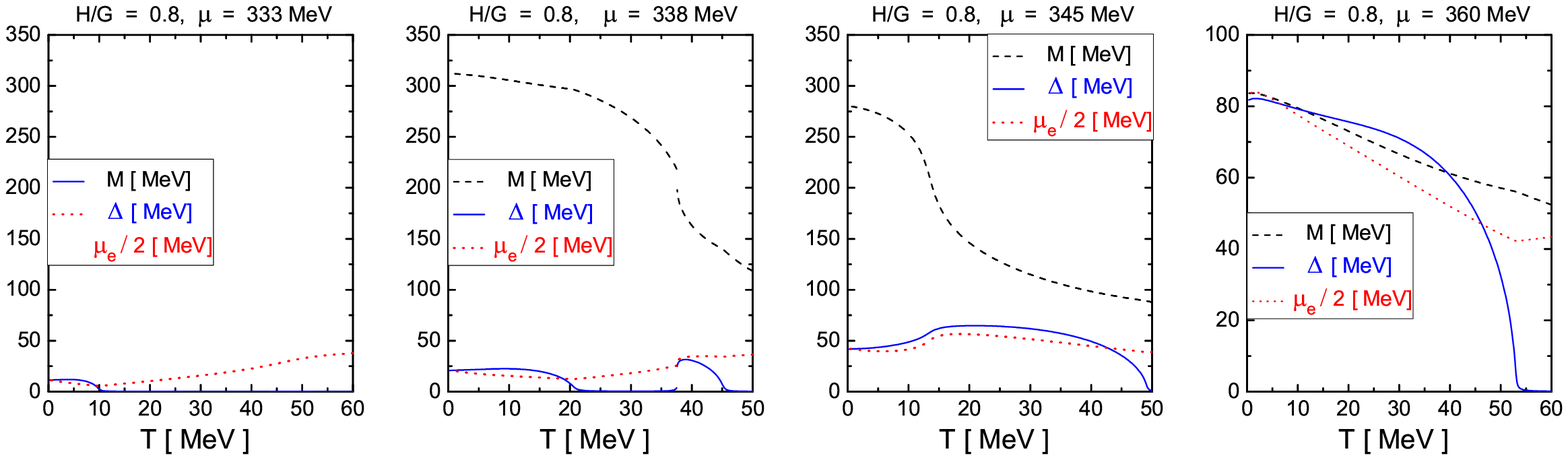}
\centerline{(b)}
\end{minipage}
\caption{$M$, $\Delta$, and $\mu_e$ as functions of the chemical
  potential in (a) and the temperature in (b) for $H/G=0.8$ under the
  condition of electric-charge neutrality.}
\label{fig:gapr80}
\end{figure}

\begin{figure}
\hspace{-.05\textwidth}
\begin{minipage}[t]{\textwidth}
\includegraphics[width=\textwidth]{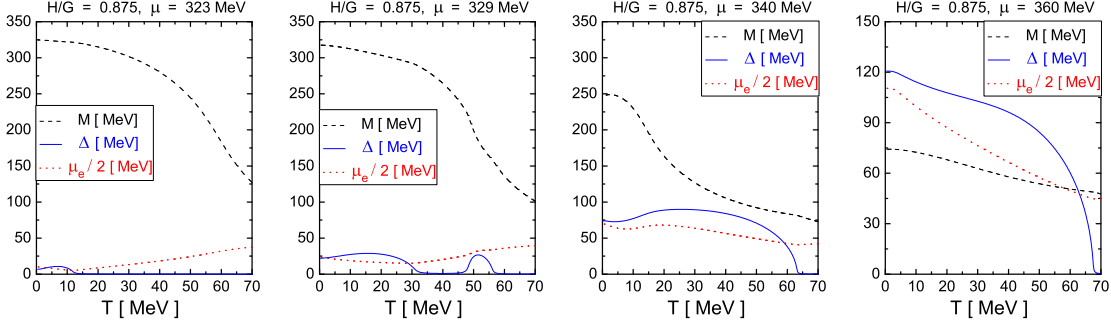}
\end{minipage}
\caption{$M$, $\Delta$, and $\mu_e$ as functions of the
  temperature for $H/G=0.875$ under the condition of electric-charge
  neutrality.}
\label{fig:gapr875}
\end{figure}

  We next choose $H/G=0.8$ and $H/G=0.875$ as examples for the
``strong'' coupling case, and present the corresponding phase diagrams
in Fig.~\ref{fig:phasestrong}.  One can see that, with increasing
diquark coupling constant, the three-critical-point structure is first
replaced by the two-critical-point structure and eventually the whole
chiral phase transition becomes crossover.  The phase boundary of the
CSC phase is always a critical line of second-order transitions, which
would be altered by gauge field fluctuations~\cite{Matsuura:2003md}.

  In contrast to Fig.~\ref{fig:medium}~(b), Fig.~\ref{fig:phasestrong}
shows that, as $H/G$ increases, the critical lines for the chiral and
diquark phase transitions shift toward the lower chemical potential
direction and the COE region is relatively enlarged.  This is a
natural result since the enhanced diquark condensate more strongly
suppresses the chiral condensate.  It is interesting that the enhanced
diquark condensate first converts the ``survivor'' line existing at
intermediate coupling into crossover as noticed from
Fig.~\ref{fig:phasestrong}~(a) and then melts the ``remnant'' line
down to crossover as shown in Fig.~\ref{fig:phasestrong}~(b).  This
means that in the COE region the diquark condensate plays a dominant
role over the chiral condensate and the original nature of the chiral
transition fails to survive at low temperature.  We remark that the
two-critical-point structure in Fig.~\ref{fig:phasestrong}~(a) is
topologically the same as what was reported in
Ref.~\cite{KitazawaVector}, which may further illustrate an analogy
between the positive $\mu_e$ and the repulsive vector interaction.

  The dependence of $M$, $\Delta$, and $\mu_e$ on $\mu$ (or $T$) for
different fixed $T$ (or $\mu$) with $H/G=0.8$ is presented in
Fig.~\ref{fig:gapr80}~(a) (or (b)).  All the figures show that all
the physical quantities behave in accord with the above picture.

  Figure~\ref{fig:gapr875} shows $M$, $\Delta$, and $\mu_e$ as
functions of $T$ at four fixed chemical potentials at $H/G=0.875$.
We find that both the remnant and the survivor are breached by the
large diquark condensate and no first-order transition remains. We
only notice that Fig.~\ref{fig:gapr875} shows again the unusual
temperature dependence of the diquark condensate, especially in
the COE region;  the diquark condensate is an increasing function
of $T$ in the low-temperature region, and then it decreases in the
higher temperature region.  The mechanism for this peculiar
behavior of the diquark condensate has been accounted for in the
last subsection.  It is worthwhile to note that the fourth panels
in Figs.~\ref{fig:gapr80}~(b) and \ref{fig:gapr875} show that, if
the diquark coupling is strong enough so that $\Delta\neq0$ at
$T=0$, the nonmonotonous behavior of the gap energy exists no
more, which only decreases with increasing temperature.

%%%   The Case of Weak Diquark Coupling   %%%

\subsection{The Case of Weak Diquark Coupling}

\begin{figure}
\hspace{-.05\textwidth}
\begin{minipage}[t]{0.4\textwidth}
\includegraphics*[width=\textwidth]{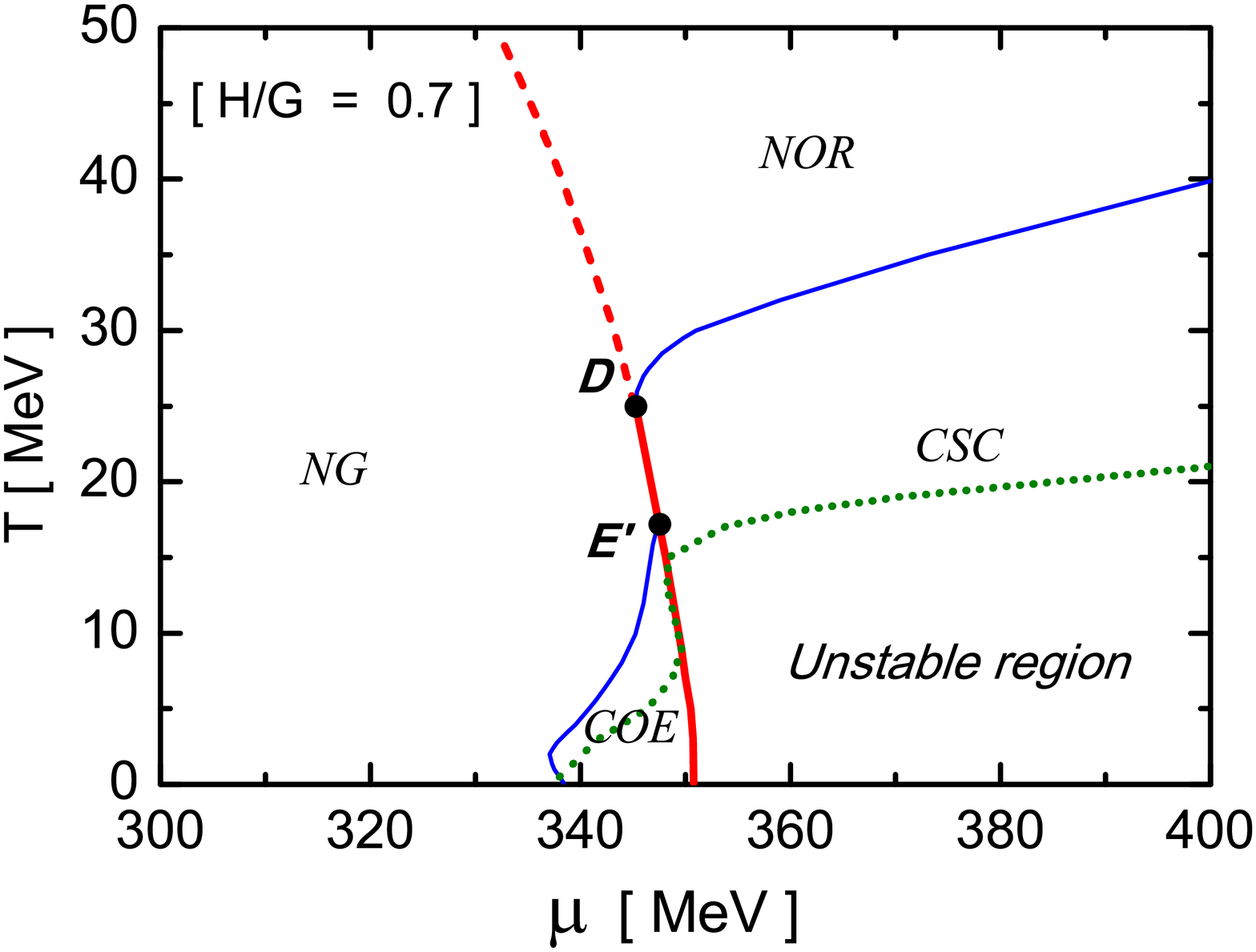}
\centerline{(a)}
\end{minipage}
\hspace{.1\textwidth}
\begin{minipage}[t]{0.4\textwidth}
\includegraphics*[width=\textwidth]{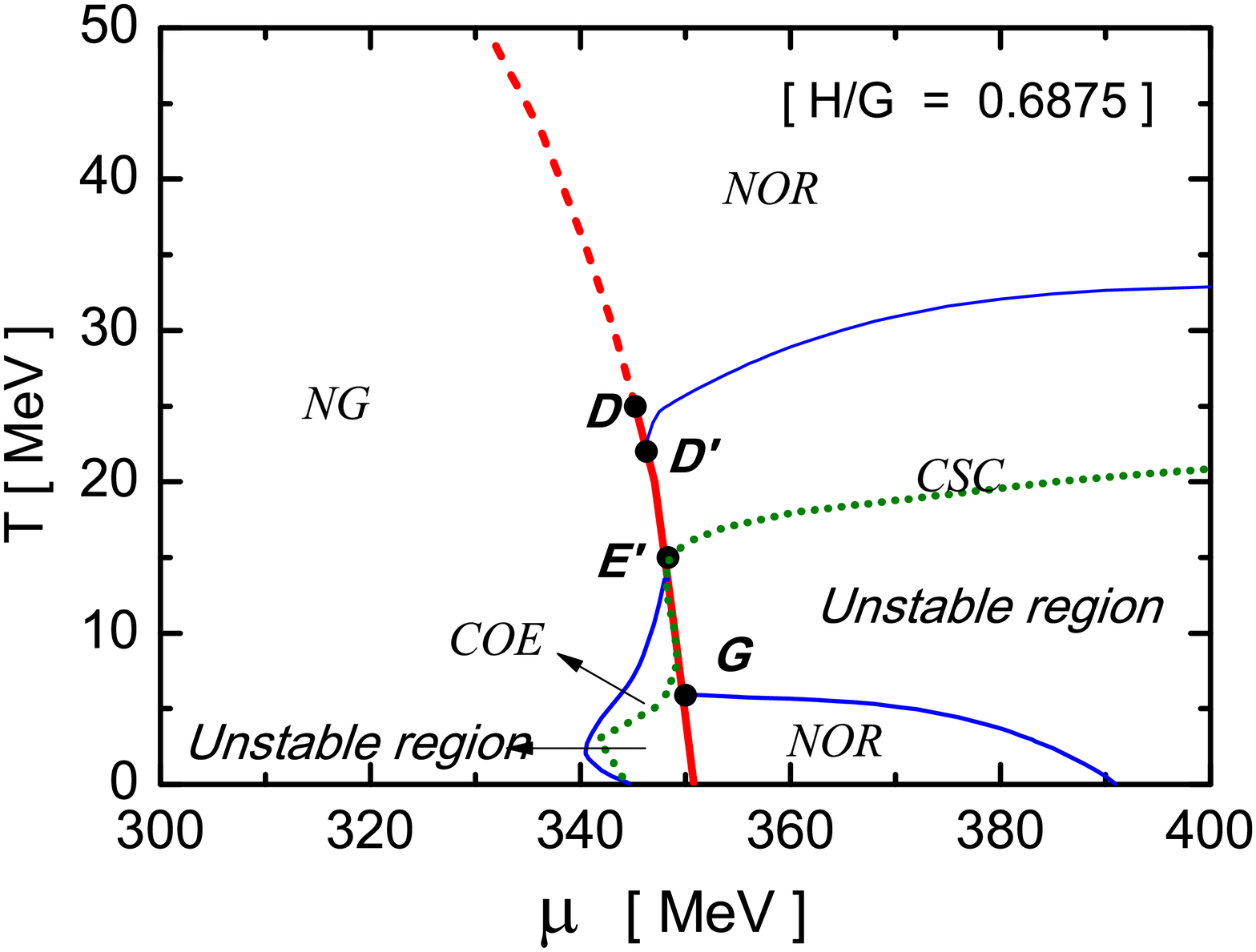}
\centerline{(b)}
\end{minipage}
\caption{The phase diagrams for the weak diquark couplings $H/G=0.7$
  and $H/G=0.6875$ under electric-charge neutrality.
 The boundary of the unstable
  region is indicated by the dotted curve as previously.}
\label{fig:phaseweak}
\end{figure}

%\begin{figure}
%\hspace{-.0\textwidth}
%\begin{minipage}[t]{0.7\textwidth}
%\includegraphics*[width=\textwidth]{gapur70.eps}
%\centerline{(a)}
%\vspace*{1mm}
%\end{minipage}
%\hspace{-.05\textwidth}
%\begin{minipage}[t]{0.99\textwidth}
%\includegraphics*[width=\textwidth]{gaptr70.eps}
%\centerline{(b)}
%\end{minipage}
%\caption{$M$, $\Delta$, and $\mu_e$ as functions of the chemical
%  potential in (a) and the temperature in (b) for $H/G=0.7$ under the
%  condition of electric-charge neutrality.}
%\label{fig:gapr70}
%\end{figure}

%\begin{figure}
%\hspace{.0\textwidth}
%\begin{minipage}[t]{\textwidth}
%\includegraphics*[width=\textwidth]{gapur6875.eps}
%\centerline{(a)} \vspace*{1mm}
%\end{minipage}
%\hspace{-.03\textwidth}
%\begin{minipage}[t]{\textwidth}
%\includegraphics*[width=\textwidth]{gaptr6875.eps}
%\centerline{(b)}
%\end{minipage}
%\caption{$M$, $\Delta$, and $\mu_e$ as functions of the chemical
%  potential in (a) and the temperature in (b) for $H/G=0.6875$ under
%  the condition of electric-charge neutrality.}
%\label{fig:gapr68}
%\end{figure}

  In this subsection, we present and discuss the phase diagrams for
``weak'' diquark coupling choosing $H/G=0.7$ and $0.6875$.  As we will
see, weakening the diquark coupling can lead to unexpected
complication in the phase diagram owing to the interplay between the
chiral and diquark correlations, which is enhanced by the
charge-neutrality constraint.

  Figure~\ref{fig:phaseweak}~(a) and (b) are the phase diagrams for
$H/G=0.7$ and $0.6875$, respectively.  Figure~\ref{fig:phaseweak}~(a)
shows that the smooth crossover line \textsf{E}-\textsf{F} seen in
Fig.~\ref{fig:medium}~(b) shrinks in a way that the critical point
\textsf{F} meets \textsf{E} at \textsf{E}' on
Fig.~\ref{fig:phaseweak}~(a).  Then, the three-critical-point
structure is replaced by the usual one-critical-point structure.  This
is because the magnitude of the diquark condensate is too small to
turn the chiral transition into crossover.
%% though the diquark
%%condensate in the COE phase remains an increasing function of $T$, as
%%long as $T$ is small, according to our numerical calculation.
%, as shown in the first panel of Fig.~\ref{fig:gapr70}~(a) and
%the first two panels of Fig.\ref{fig:gapr68}~(a).
%%Here it is noteworthy that this unconventional temperature
%%dependence of the diquark condensate persisting even in the CSC
%%phase in the weak coupling case. %, as shown in the fourth panel of
%Fig.~\ref{fig:gapr70}~(b).

Let us briefly discuss a notable point that, for $H/G=0.6875$, the
critical point \textsf{D} and the triple point denoted by
\textsf{D}$'$ can become separate.  It seems that, as long as the
diquark coupling is larger than a certain critical value, there is a
mechanism to make \textsf{D} and \textsf{D}$'$ coincide.  A similar
observation is reported in the $2+1$ flavor case too~\cite{Abuki}.  If
$H/G$ is lowered further, our numerical calculation results in the
shrinking COE and the enhancing (low-$T$) NOR regions.

Although we can recognize an intriguing structure in
Fig.~\ref{fig:phaseweak}~(b) in comparison to (a) -- the emergence
of a NOR ``island'' surrounded by the CSC phase at low $T$ -- we
will not take a close look into this region.  This is because the
newly arising NOR phase and the triple point \textsf{G} in
Fig.~\ref{fig:phaseweak}~(b) are located deeply in the unstable
region, the boundary of which is  indicated by the dotted curve.
  We shall address the
instability problem in the next subsection.

\subsection{Chromomagnetic Instability}
\label{sec:Meissner}

\begin{figure}
%\hspace{-.0\textwidth}
%\begin{minipage}[t]{0.23\textwidth}
%\includegraphics*[width=\textwidth]{comparet05.eps}
%\centerline{(a)}
%\end{minipage}
%\hspace{.0\textwidth}
\begin{minipage}[t]{0.3\textwidth}
\includegraphics*[width=\textwidth]{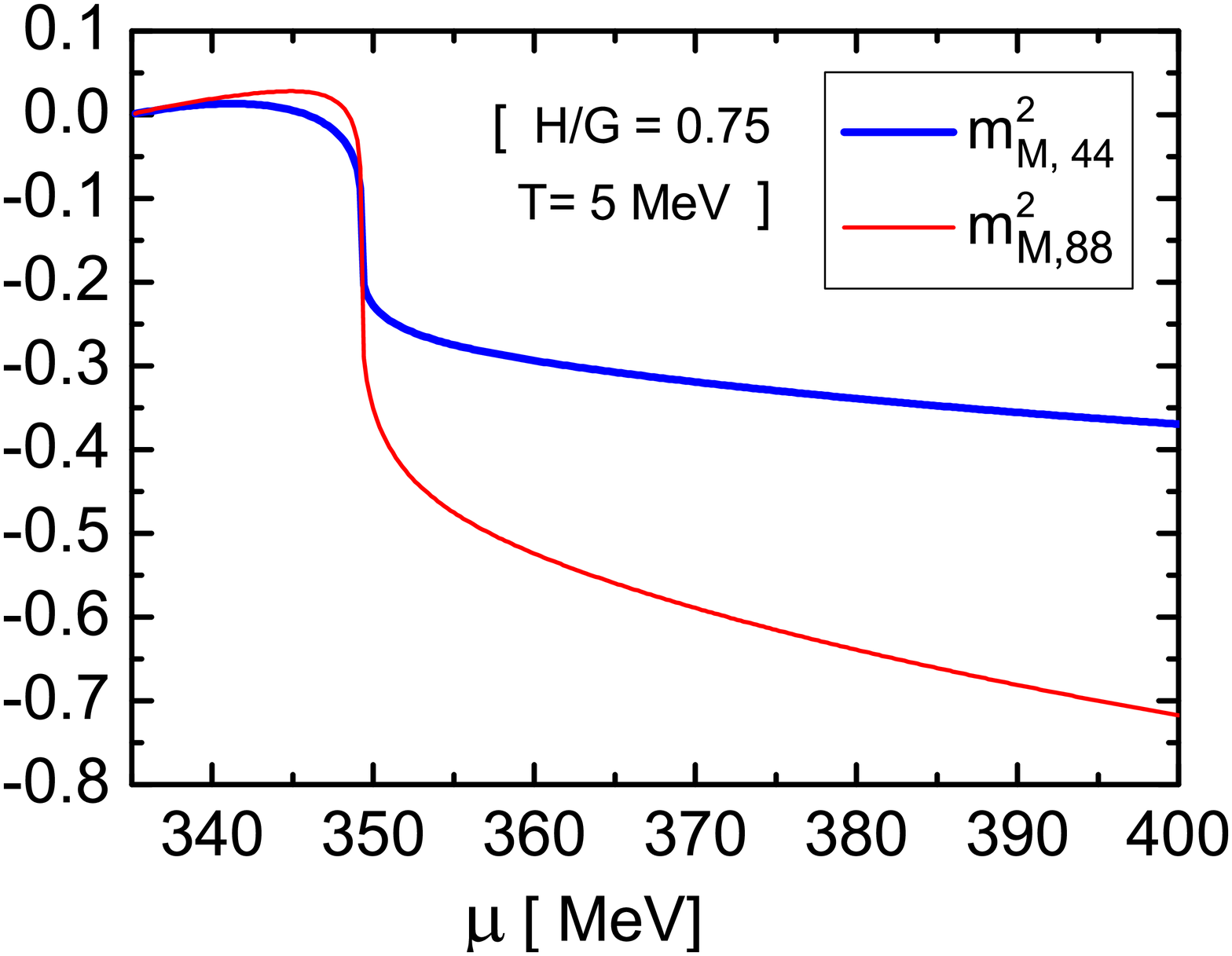}
\centerline{(a)}
\end{minipage}
\hspace{.0\textwidth}
\begin{minipage}[t]{0.3\textwidth}
\includegraphics*[width=\textwidth]{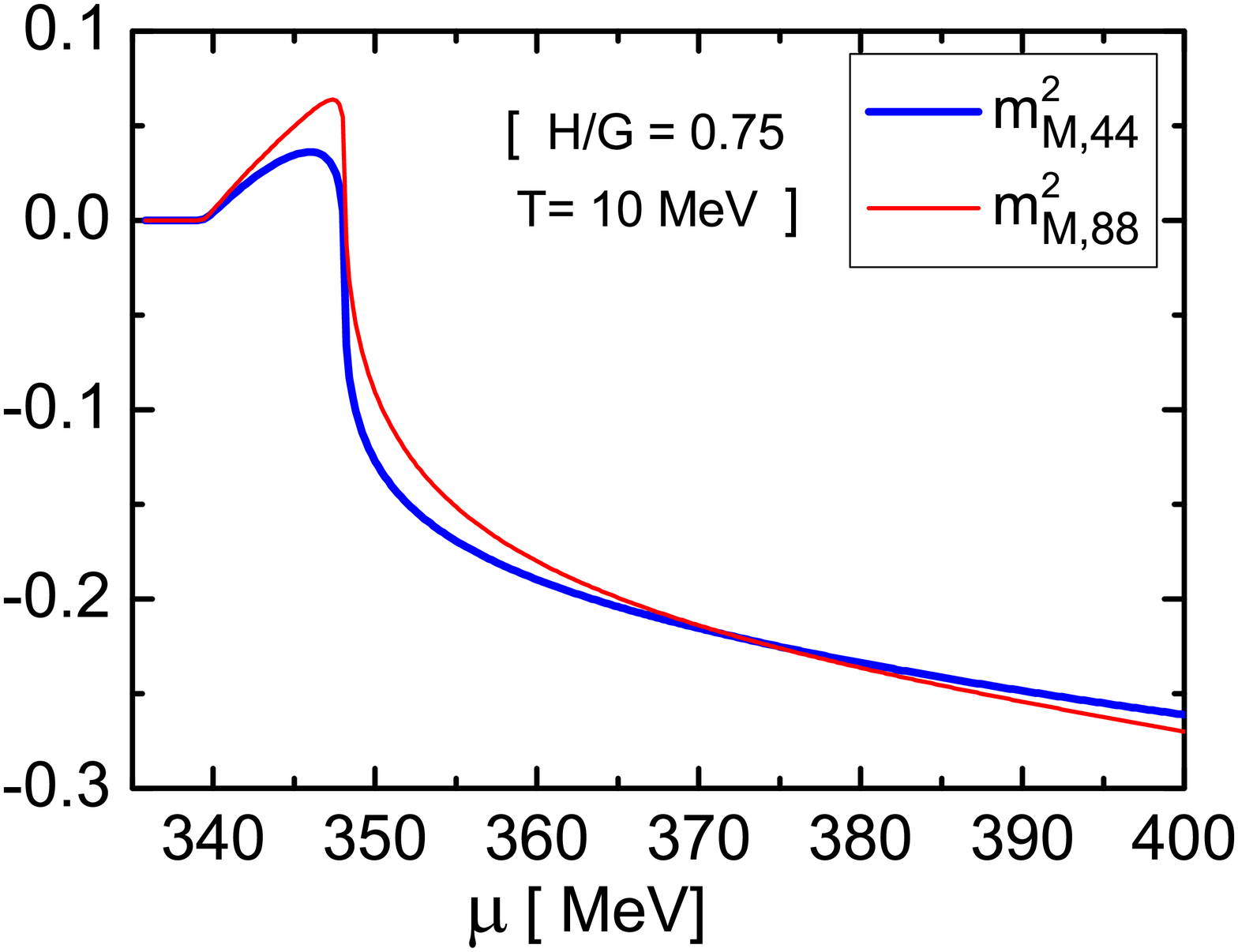}
\centerline{(b)}
\end{minipage}
\hspace{.0\textwidth}
\begin{minipage}[t]{0.3\textwidth}
\includegraphics*[width=\textwidth]{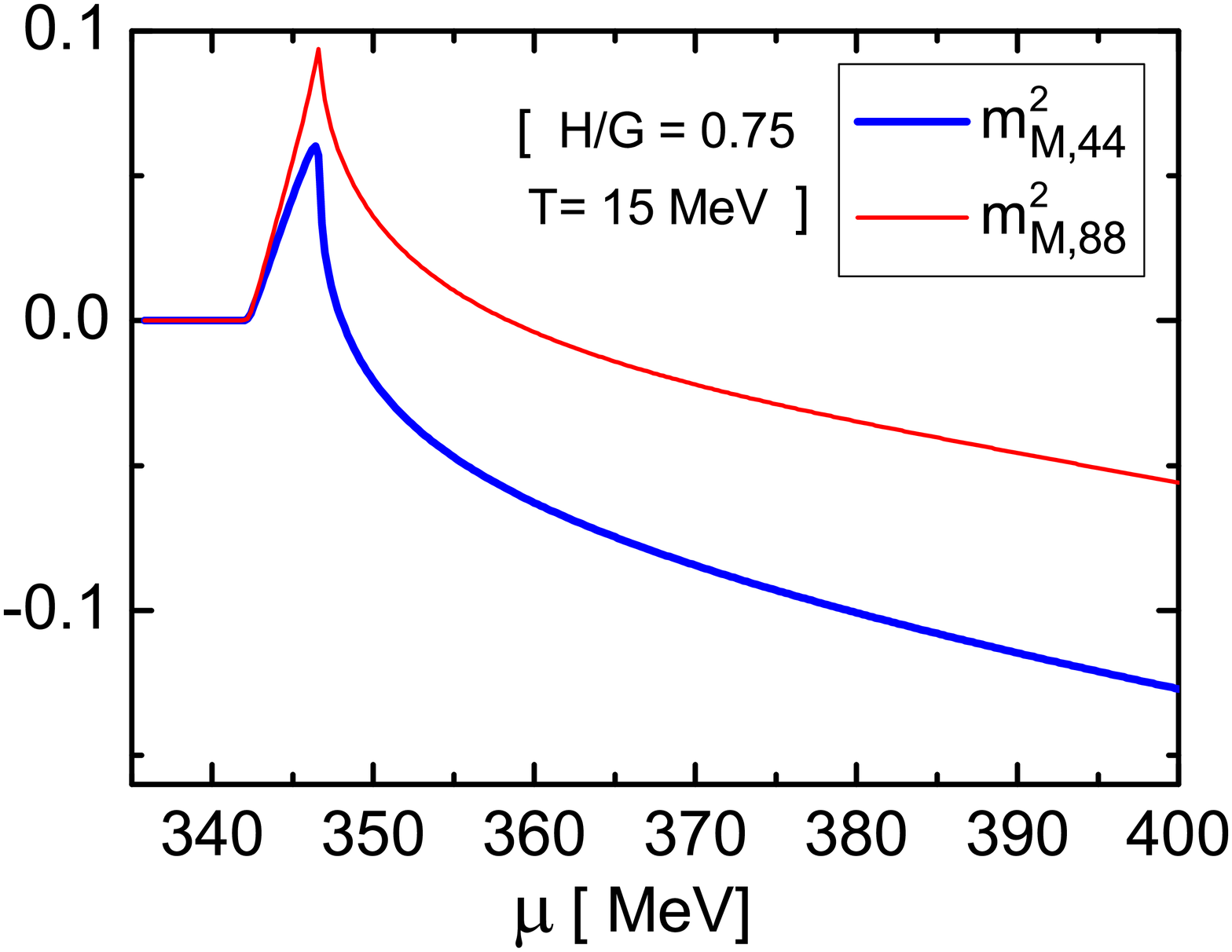}
\centerline{(c)}
\end{minipage}
\caption{Meissner masses squared as functions of $\mu$ for
different temperatures with $H/G=0.75$. All the results are in the
unit $m_g^2=4\alpha_s^2\bar{\mu}^2/(3\pi)$.}
\label{fig:meissnermass}
\end{figure}

It has been known that the homogeneous CSC state could suffer from
instability~\cite{Huang:2004bg,Alford:2005qw,Casalbuoni:2004tb,Fukushima:2005cm,
Kiriyama:2006jp} when the Fermi surface mismatch grows comparable
to the pairing gap. The instability occurs in various channels
simultaneously~\cite{Iida:2006df}, among which the transverse
gluon field triggers the chromomagnetic instability.  We can
theoretically perceive the instability by the negative Meissner
screening mass squared (or pure imaginary Meissner mass).  In
Figs.~\ref{fig:medium}~(b), \ref{fig:phasestrong}, and
\ref{fig:phaseweak} we have located the unstable region  given by
the condition that either of eight gluons has negative Meissner
mass squared.  In what follows let us first explain how we
calculate the Meissner mass.

In the two-flavor case the analytical expressions for the Meissner
mass are available in the hard dense loop (HDL)
approximation~\cite{Huang:2004bg,He:2007cn}.  In our evaluation we
have utilized them with our model parameters substituted to the
formulae.  We draw the dotted curves in
Figs.~\ref{fig:medium}~(b), \ref{fig:phasestrong}, and
\ref{fig:phaseweak} in this way.  At the same time we have carried
out the brute-force computation by means of the potential
curvature with respect to the gauge field source
\cite{Kiriyama:2006jp}.
 This
calculation is necessary for the numerical check of
the consistency  between the HDL in Ref.~\cite{He:2007cn}
and the approximation made in this work.
  We have in fact confirmed the consistency in the case that we
keep using a sharp three-momentum cutoff $\Lambda=651\MeV$.  It should
be mentioned that how to renormalize the unphysical cutoff dependence
is not yet known in a field-theory manner~\cite{Alford:2005qw}.  Our
prescription is motivated in accord with the NJL model treatment,
while a larger $\Lambda$ would make the unstable regions smaller.  In
this sense, even though the dotted curves in
Figs.~\ref{fig:medium}~(b), \ref{fig:phasestrong}, \ref{fig:phaseweak}
might move downward, the cutoff dependence would not affect the stable
regions above them.

We show the Meissner masses squared for the fourth and eighth
gluons at $H/G=0.75$ as functions of $\mu$ in
Fig.~\ref{fig:meissnermass}. We note that the Meissner mass
squared changes rapidly from NG to COE leading to the
chromomagnetic instability in the high-$\mu$ and low-$T$ region of
COE and CSC as indicated by the dotted curve in
Fig.~\ref{fig:medium}~(b).

Figures~\ref{fig:medium}~(b) and \ref{fig:phasestrong}~(a)
 tell us  that two critical points associated with the
\textsf{D}-\textsf{E} line are outside the unstable region.
 However, the instability analysis solely can not
 tell
% Regarding
the fate of  another critical point \textsf{F} in reality.
% or not is unclear from the instability analysis.
 In fact,
% Here we would point out
there are two possibilities to accommodate three critical points;
  one is that the
unstable region may shrink to lower temperature if the cutoff is
taken large and the other is that some other $H/G$ value may push
\textsf{F} up above the unstable region. The latter case can been
confirmed in the current model study when $H/G$ is located in a
narrow region near but below 0.75. To settle the robustness of
\textsf{F}, we have to go into the identification of the true
ground state inside the unstable region.

There are some attempts to overcome the chromomagnetic
instability. The instability has a favorite direction leading to
the (colored) Larkin-Ovchinnikov-Fulde-Ferrell (LOFF)
state~\cite{Giannakis:2004pf,Iida:2006df,Fukushima:2006su} (or
spontaneous current generation~\cite{Huang:2005pv,Schafer:2005ym},
which shares the same mathematical structure with the colored
plane-wave LOFF description~\cite{Fukushima:2005fh}).  We note
that the gluonic phase~\cite{Gorbar:2005rx} with one gluon
condensate can translate into the plane-wave LOFF state, while a
more stable gluonic phase with multiple gluon condensates has a
chance to surpass the LOFF state~\cite{Kiriyama:2006ui}.  The
stable-unstable boundary we drew on the phase diagram can be
regarded as a second-order phase transition line from the
homogeneous CSC phase to one of these inhomogeneous states.
Therefore, even though \textsf{F} in Fig.~\ref{fig:medium}~(b) is
overridden by the instability or the second-order phase
transition, the physical consequence is very similar to the
three-critical point structure, that is, one of the critical
points is replaced by the critical line.  Interestingly enough,
then, Fig.~\ref{fig:phasestrong}~(a) is indistinguishable from
Fig.~\ref{fig:medium}, meaning that the three-critical point (or
line) interpretation is in effect elongated for a wider range of
$H/G$ due to the chromomagnetic instability.

%%%   Discussions   %%%

\subsection{Discussions}

  In the previous subsections, we have given a detailed account for
four possible scenarios for the critical point structure with the
influence of the diquark condensate under the electric-charge
neutrality.  For the model parameters adopted here, numerical
calculations suggest that the three-critical-point structure of the
phase diagram appears in the range $0.735\lesssim H/G\lesssim 0.767$,
while the two-critical-point structure in the range
$0.767\lesssim H/G\lesssim 0.82$.  For the stronger coupling region
$H/G\gtrsim0.82$, the chiral phase transition remains crossover, while
for the weaker coupling region $H/G\lesssim0.735$, only
one-critical-point structure appears.

  We have checked that the region with multi-critical-point structure
is slightly modified quantitatively when nonzero $\mu_8$ is taken into
account;  the range $0.745\lesssim H/G\lesssim 0.772$ corresponds to
the three-critical-point structure and
$0.772\lesssim H/G\lesssim 0.80$ to the two-critical-point one.  Thus
we can conclude that our main results on the novel phase structures
are not altered even when the color-charge neutrality is fully
incorporated although the parameter range leading to the
multi-critical-point structure may be somewhat narrowed.

  It should be noted that the multi-critical-point structure was not
pointed out in the previous works such as Ref.~\cite{Abuki} where
the electric neutrality and the dynamical quark mass were
simultaneously considered for $2+1$ flavor quark matter.  Since
the strange quark mass is still as large as the quark chemical
potential near the chiral phase transition, the $u$ and $d$ quarks
should play the dominant role in the vicinity of this region.
Then, what causes the difference in the phase diagram?  We have
found that the appearance of the multi-critical-point structure is
sensitive to the value of the constituent quark mass in the
vacuum.  In the present study the vacuum constituent quark mass
for light quarks is $325.5\MeV$, while it is $400\MeV$ in
Ref.~\cite{Abuki}.  Notice that the larger the constituent quark
mass is, the smaller the Fermi sphere becomes, disfavoring the
diquark pairing for a given $\mu$.  Therefore, in
Ref.~\cite{Abuki}, the strong first-order chiral transition at low
temperature is hardly affected by the diquark even with $\mu_e$,
which could make the chiral transition smooth.  However, in turn,
it suggests the possibility that the multi-critical-point
structure may appear even with the large constituent quark mass
adopted in Ref.~\cite{Abuki} if the repulsive vector interaction
is also included, which is rather close to the realistic
situation.

%%%%%%%%%%   CONCLUSIONS AND OUTLOOK   %%%%%%%%%%

\section{CONCLUSIONS AND OUTLOOK}

  In this paper, we have explored the effect of electric-charge
neutrality in $\beta$ equilibrium on the chiral phase transition
with and without considering the diquark condensate within a
simple two-flavor NJL model.

  We first disclosed the similarity of the roles on the chiral phase
transition played by the positive electric chemical potential and
the repulsive vector interaction as follows:  1) Both of them make
the chiral critical line at low temperature shift toward larger
quark chemical potential and effectively weaken the first-order
chiral phase transition.  2) Both of them effectively enhance the
competition between the chiral and diquark condensates.

  In some model parameter region including the diquark
condensate, the combination of these two properties can result in
the emergence of the two-critical-point structure for the chiral
phase transition in the electric-charge neutral case.  It is
noticeable that the magnitude of the repulsive vector interaction
in Ref.~\cite{KitazawaVector} was an assumption although the
existence itself of such an interaction has a generic foundation;
on the other hand, the positive electric chemical potential in
this paper is self-consistently determined by the physical
requirement that the bulk matter must be under the
charge-neutrality condition.

  Besides two nontrivial effects mentioned above, our investigation
also showed a quite unconventional three-critical-point structure
for the chiral phase transition.  This result is directly
associated with the abnormal behavior that the diquark energy gap
can increase with increasing temperature in the coexisting region
for a certain range of $T$-$\mu$ and $H/G$, which is possible
under the tangled influences of electric-charge neutrality and the
interplay between the chiral and diquark condensates.  To our
knowledge, so far, this is the first case which concretely
demonstrates the three-critical-point structure for the chiral
phase transition.  Although this result with three critical points
could be sensitive to the model parameter choice, it definitely
gives a hint that there may exist a possibility that the
complexity in QCD allows for more complicated phase boundaries
associated with the chiral phase transition.  Moreover we  have
studied the chromomagnetic instability and found that the
three-critical-point scenario may be taken over by the structure
with two critical points and one critical line.  Our result also
has a meaningful implication for the study of phase transitions
leading to a coexisting phase involving superconductivity in
condensed matter physics when external conditions are enforced to
the system like the neutrality constraint causing the mismatched
Fermi surfaces.

  In view of the common effects on the chiral phase transition by the
positive $\mu_e$ and the repulsive vector interaction $G_{\rm V}$,
it is natural to expect that the influences mentioned above will
be greatly enhanced when these two aspects are considered
simultaneously.  It is expected that the three-critical-point
structure may extend to a weaker coupling region in the presence
of the vector interaction, although the number of the critical
points might change in a different way from that depicted in
Fig.~\ref{fig:schematic} when the vector coupling is increased.

  In addition, the confinement-deconfinement phase transition at finite
temperature and density is still poorly understood inside the CSC
phase, that is because of the lack of confinement in the NJL-type
model.  Recently the PNJL was proposed~\cite{Polyakov,PlApp1} and
had been extensively used to study the thermal properties and the
phase transition of
QCD~\cite{Roessner:2006xn,Megias:2004hj,PlApp2,Zhang:2007,Hansen:2006ee,PlApp4,PlApp3,PlApp5,PlApp6,PlApp8}.
It is interesting to investigate whether the results obtained in
this paper are also true when including the Polyakov loop
dynamics. Practically, the PNJL model tends to push the location
of the critical point up toward the higher temperature region.
That means, a detailed structure observed in this work could be
magnified by the Polyakov loop.  More details on these two issues
will be reported in our future publication~\cite{future}, and in
fact, we have already verified that the multi-critical-point
structure still exists when the Polyakov loop effect is taken into
account.

\acknowledgments

  One of the authors Z.~Zhang thanks the support of the Grants-in-aid
provided by the Japan Society for the Promotion of Science (JSPS).
This work was partially supported by a Grant-in-Aid for Scientific
Research by the Ministry of Education, Culture, Sports, Science
and Technology (MEXT) of Japan (No. 20540265, No.
20740134,No.19$\cdot$07797),
 by the Yukawa International Program for Quark-Hadron Sciences, and by the
Grant-in-Aid for the global COE program `` The Next Generation of
Physics, Spun from Universality and Emergence '' from MEXT.


\begin{thebibliography}{99}

\bibitem{WilczekReview}
% \bibitem{Rajagopal:2000wf}
  K.~Rajagopal and F.~Wilczek,
  %``The condensed matter physics of QCD,''
  arXiv:hep-ph/0011333.

\bibitem{RischkeReview}
%\bibitem{Rischke:2003mt}
  D.~H.~Rischke,
  %``The quark-gluon plasma in equilibrium,''
  Prog.\ Part.\ Nucl.\ Phys.\  {\bf 52}, 197 (2004)
  [arXiv:nucl-th/0305030].

\bibitem{BuballaReview}
%\bibitem{Buballa:2003qv}
  M.~Buballa,
  %``NJL model analysis of quark matter at large density,''
  Phys.\ Rept.\  {\bf 407}, 205 (2005)
  [arXiv:hep-ph/0402234].

\bibitem{AlfordReview}
%\bibitem{Alford:2007xm}
  M.~G.~Alford, A.~Schmitt, K.~Rajagopal and T.~Schafer,
  %``Color superconductivity in dense quark matter,''
  arXiv:0709.4635 [hep-ph].

\bibitem{CFL}
%\bibitem{Alford:1998mk}
  M.~G.~Alford, K.~Rajagopal and F.~Wilczek,
  %``Color-flavor locking and chiral symmetry breaking in high density {QCD},''
  Nucl.\ Phys.\  B {\bf 537}, 443 (1999)
  [arXiv:hep-ph/9804403].

\bibitem{Hatsuda:2006ps}
  T.~Hatsuda, M.~Tachibana, N.~Yamamoto and G.~Baym,
  %``New critical point induced by the axial anomaly in dense QCD,''
  Phys.\ Rev.\ Lett.\  {\bf 97}, 122001 (2006)
  [arXiv:hep-ph/0605018];
%\bibitem{Yamamoto:2007ah}
%  N.~Yamamoto, M.~Tachibana, T.~Hatsuda and G.~Baym,
  %``Phase structure, collective modes, and the axial anomaly in dense QCD,''
  Phys.\ Rev.\  D {\bf 76}, 074001 (2007)
  [arXiv:0704.2654 [hep-ph]].

\bibitem{Continuity}
%\bibitem{Schafer:1998ef}
  T.~Schafer and F.~Wilczek,
  %``Continuity of quark and hadron matter,''
  Phys.\ Rev.\ Lett.\  {\bf 82}, 3956 (1999)
  [arXiv:hep-ph/9811473];
%\bibitem{Alford:1999pa}
  M.~G.~Alford, J.~Berges and K.~Rajagopal,
  %``Unlocking color and flavor in superconducting strange quark matter,''
  Nucl.\ Phys.\  B {\bf 558}, 219 (1999)
  [arXiv:hep-ph/9903502].

\bibitem{KitazawaVector}
%\bibitem{Kitazawa:2002bc}
  M.~Kitazawa, T.~Koide, T.~Kunihiro and Y.~Nemoto,
  %``Chiral and color superconducting phase transitions with vector interaction
  %in a simple model,''
  Prog.\ Theor.\ Phys.\  {\bf 108}, 929 (2002)
  [arXiv:hep-ph/0207255].

\bibitem{Asakawa}
%\bibitem{Asakawa:1989bq}
  M.~Asakawa and K.~Yazaki,
  %``CHIRAL RESTORATION AT FINITE DENSITY AND TEMPERATURE,''
  Nucl.\ Phys.\  A {\bf 504}, 668 (1989).

\bibitem{Klimt}
%\bibitem{Klimt:1990ws}
  S.~Klimt, M.~Lutz and W.~Weise,
  %``Chiral phase transition in the SU(3) Nambu and Jona-Lasinio model,''
  Phys.\ Lett.\  B {\bf 249}, 386 (1990).

\bibitem{Buballa1996}
%\bibitem{Buballa:1996tm}
  M.~Buballa,
  %``The problem of matter stability in the Nambu-Jona-Lasinio model,''
  Nucl.\ Phys.\  A {\bf 611}, 393 (1996)
  [arXiv:nucl-th/9609044].

\bibitem{Lattice2color}
%\bibitem{Kogut:2002cm}
  J.~B.~Kogut, D.~Toublan and D.~K.~Sinclair,
  %``The phase diagram of four flavor SU(2) lattice gauge theory at nonzero
  %chemical potential and temperature,''
  Nucl.\ Phys.\  B {\bf 642}, 181 (2002)
  [arXiv:hep-lat/0205019].

\bibitem{Klevansky:1992}
%\bibitem{Klevansky:1992qe}
  S.~P.~Klevansky,
  %``The Nambu-Jona-Lasinio model of quantum chromodynamics,''
  Rev.\ Mod.\ Phys.\  {\bf 64}, 649 (1992).

\bibitem{Hatsuda:1994}
%\bibitem{Hatsuda:1994pi}
  T.~Hatsuda and T.~Kunihiro,
  %``QCD phenomenology based on a chiral effective Lagrangian,''
  Phys.\ Rept.\  {\bf 247}, 221 (1994)
  [arXiv:hep-ph/9401310].

\bibitem{future}
 Z.~Zhang and T.~Kunihiro, work under completion.

\bibitem{Hatsuda:1985ey}
  T.~Hatsuda and T.~Kunihiro,
  %``Critical Phenomena Associated With Chiral Symmetry Breaking And Restoration
  %In QCD,''
  Prog.\ Theor.\ Phys.\  {\bf 74} (1985) 765.

\bibitem{Zhang:2007}
%\bibitem{Zhang:2006gu}
  Z.~Zhang and Y.~X.~Liu,
  %``Coupling of pion condensate, chiral condensate and Polyakov loop in an
  %extended NJL model,''
  Phys.\ Rev.\  C {\bf 75}, 064910 (2007)
  [arXiv:hep-ph/0610221].

\bibitem{Alford:2002kj}
  M.~Alford and K.~Rajagopal,
  %``Absence of two-flavor color superconductivity in compact stars,''
  JHEP {\bf 0206}, 031 (2002)
  [arXiv:hep-ph/0204001];
%\bibitem{Steiner:2002gx}
  A.~W.~Steiner, S.~Reddy and M.~Prakash,
  %``Color-neutral superconducting quark matter,''
  Phys.\ Rev.\  D {\bf 66}, 094007 (2002)
  [arXiv:hep-ph/0205201].

\bibitem{Roessner:2006xn}
  S.~Roessner, C.~Ratti and W.~Weise,
  %``Polyakov loop, diquarks and the two-flavour phase diagram,''
  Phys.\ Rev.\  D {\bf 75}, 034007 (2007)
  [arXiv:hep-ph/0609281].

\bibitem{PlApp8}
%\bibitem{Abuki:2008ht}
  H.~Abuki, M.~Ciminale, R.~Gatto, G.~Nardulli and M.~Ruggieri,
  %``Enforced neutrality and color-flavor unlocking in the three-flavor
  %Polyakov-loop NJL model,''
  Phys.\ Rev.\  D {\bf 77}, 074018 (2008)
  [arXiv:0802.2396 [hep-ph]];
%\bibitem{Abuki:2008nm}
  H.~Abuki, R.~Anglani, R.~Gatto, G.~Nardulli and M.~Ruggieri,
  %``Chiral crossover, deconfinement and quarkyonic matter within a Nambu-Jona
  %Lasinio model with the Polyakov loop,''
  arXiv:0805.1509 [hep-ph].

\bibitem{Huang}
%\bibitem{Shovkovy:2003uu}
  M.~Huang and I.~Shovkovy,
%  I.~Shovkovy and M.~Huang,
  %``Gapless two-flavor color superconductor,''
  Phys.\ Lett.\  B {\bf 564}, 205 (2003)
  [arXiv:hep-ph/0302142];
%\bibitem{Huang:2003xd}
%  M.~Huang and I.~Shovkovy,
  %``Gapless color superconductivity at zero and at finite temperature,''
  Nucl.\ Phys.\  A {\bf 729}, 835 (2003)
  [arXiv:hep-ph/0307273].

\bibitem{foot:1}
When the diquark condensate and the charge-neutrality constraint
are not taken into consideration, the $\mathrm{U_A}(1)$ anomaly term
as given by the t'~Hooft determinantal interaction in the three-flavor
case does not give rise to any difference between the constituent
quark masses once the current quark masses are set equal, as shown in
Ref.~\cite{Kunihiro:1989my}; see also Ref.~\cite{Hatsuda:1994}.

\bibitem{Kunihiro:1989my}
  T.~Kunihiro,
  %``Effects Of The U(A)(1) Anomaly On The Quark Condensates And Meson
  %Properties At Finite Temperature,''
  Phys.\ Lett.\  B {\bf 219} (1989) 363.

\bibitem{Rapp}
%\bibitem{Rapp:1999qa}
  R.~Rapp, T.~Schafer, E.~V.~Shuryak and M.~Velkovsky,
  %``High-density QCD and instantons,''
  Annals Phys.\  {\bf 280}, 35 (2000)
  [arXiv:hep-ph/9904353].

\bibitem{Carter}
%\bibitem{Carter:1998ji}
  G.~W.~Carter and D.~Diakonov,
  %``Light quarks in the instanton vacuum at finite baryon density,''
  Phys.\ Rev.\  D {\bf 60}, 016004 (1999)
  [arXiv:hep-ph/9812445].

\bibitem{RadomMatrix}
%\bibitem{Vanderheyden:2000ti}
  B.~Vanderheyden and A.~D.~Jackson,
  %``Random matrix model for chiral symmetry breaking and color
  %superconductivity in QCD at finite density,''
  Phys.\ Rev.\  D {\bf 62}, 094010 (2000)
  [arXiv:hep-ph/0003150].

\bibitem{Matsuura:2003md}
  T.~Matsuura, K.~Iida, T.~Hatsuda and G.~Baym,
  %``Thermal fluctuations of gauge fields and first order phase transitions  in
  %color superconductivity,''
  Phys.\ Rev.\  D {\bf 69}, 074012 (2004)
  [arXiv:hep-ph/0312042];
%\bibitem{Giannakis:2004xt}
  I.~Giannakis, D.~f.~Hou, H.~c.~Ren and D.~H.~Rischke,
  %``Gauge field fluctuations and first-order phase transition in color
  %superconductivity,''
  Phys.\ Rev.\ Lett.\  {\bf 93}, 232301 (2004)
  [arXiv:hep-ph/0406031].

\bibitem{Abuki}
%\bibitem{Abuki:2005ms}
  H.~Abuki and T.~Kunihiro,
  %``Extensive study of phase diagram for charge neutral homogeneous quark
  %matter affected by dynamical chiral condensation: Unified picture for
  %thermal unpairing transitions from weak to strong coupling,''
  Nucl.\ Phys.\  A {\bf 768}, 118 (2006)
  [arXiv:hep-ph/0509172].

\bibitem{Huang:2004bg}
  M.~Huang and I.~A.~Shovkovy,
  %``Chromomagnetic instability in dense quark matter,''
  Phys.\ Rev.\  D {\bf 70}, 051501(R) (2004)
  [arXiv:hep-ph/0407049];
%\bibitem{Huang:2004am}
%  M.~Huang and I.~A.~Shovkovy,
  %``Screening masses in neutral two-flavor color superconductor,''
%  Phys.\ Rev.\  D {\bf 70}, 094030 (2004)
  \textit{ibid.} D {\bf 70}, 094030 (2004)
  [arXiv:hep-ph/0408268].

\bibitem{Alford:2005qw}
  M.~Alford and Q.~h.~Wang,
  %``Photons in gapless color-flavor-locked quark matter,''
  J.\ Phys.\ G {\bf 31}, 719 (2005)
  [arXiv:hep-ph/0501078].

\bibitem{Casalbuoni:2004tb}
  R.~Casalbuoni, R.~Gatto, M.~Mannarelli, G.~Nardulli and M.~Ruggieri,
  %``Meissner masses in the gCFL phase of QCD,''
  Phys.\ Lett.\  B {\bf 605}, 362 (2005)
  [Erratum-ibid.\  B {\bf 615}, 297 (2005)]
  [arXiv:hep-ph/0410401].

\bibitem{Fukushima:2005cm}
  K.~Fukushima,
  %``Analytical and numerical evaluation of the Debye and Meissner masses in
  %dense neutral three-flavor quark matter,''
  Phys.\ Rev.\  D {\bf 72}, 074002 (2005)
  [arXiv:hep-ph/0506080].

\bibitem{Kiriyama:2006jp}
  O.~Kiriyama,
  %``Chromomagnetic instability in two-flavor quark matter at nonzero
  %temperature,''
  Phys.\ Rev.\  D {\bf 74}, 114011 (2006)
  [arXiv:hep-ph/0609185].

\bibitem{Iida:2006df}
  K.~Iida and K.~Fukushima,
  %``Instability of a gapless color superconductor with respect to
  %inhomogeneous fluctuations,''
  Phys.\ Rev.\  D {\bf 74}, 074020 (2006)
  [arXiv:hep-ph/0603179].

\bibitem{He:2007cn}
L.~He, M.~Jin and P.~Zhuang,
  %``Neutral color superconductivity including inhomogeneous phases at  finite
  %temperature,''
  Phys.\ Rev.\  D {\bf 75}, 036003 (2007)
  [arXiv:hep-ph/0610121].

\bibitem{Giannakis:2004pf}
  I.~Giannakis and H.~C.~Ren,
  %``Chromomagnetic instability and the LOFF state in a two flavor color
  %superconductor,''
  Phys.\ Lett.\  B {\bf 611}, 137 (2005)
  [arXiv:hep-ph/0412015].

\bibitem{Fukushima:2006su}
  K.~Fukushima,
  %``Characterizing the Larkin-Ovchinnikov-Fulde-Ferrel phase induced by the
  %chromomagnetic instability,''
  Phys.\ Rev.\  D {\bf 73}, 094016 (2006)
  [arXiv:hep-ph/0603216].

\bibitem{Huang:2005pv}
  M.~Huang,
  %``Spontaneous current generation in the 2SC phase,''
  Phys.\ Rev.\  D {\bf 73}, 045007 (2006)
  [arXiv:hep-ph/0504235].

\bibitem{Schafer:2005ym}
  T.~Schafer,
  %``P-wave meson condensation in high density QCD,''
  Phys.\ Rev.\ Lett.\  {\bf 96}, 012305 (2006)
  [arXiv:hep-ph/0508190].

\bibitem{Fukushima:2005fh}
  K.~Fukushima,
  %``Phase structure and instability problem in color superconductivity,''
  arXiv:hep-ph/0510299.

\bibitem{Gorbar:2005rx}
  E.~V.~Gorbar, M.~Hashimoto and V.~A.~Miransky,
  %``Gluonic phase in neutral two-flavor dense QCD,''
  Phys.\ Lett.\  B {\bf 632}, 305 (2006)
  [arXiv:hep-ph/0507303];
%\bibitem{Gorbar:2005tx}
%  E.~V.~Gorbar, M.~Hashimoto and V.~A.~Miransky,
  %``Neutral LOFF state and chromomagnetic instability in two-flavor dense
  %QCD,''
  Phys.\ Rev.\ Lett.\  {\bf 96}, 022005 (2006)
  [arXiv:hep-ph/0509334];
%\bibitem{Gorbar:2007vx}
%  E.~V.~Gorbar, M.~Hashimoto and V.~A.~Miransky,
  %``Gluonic phases, vector condensates, and exotic hadrons in dense QCD,''
  Phys.\ Rev.\  D {\bf 75}, 085012 (2007)
  [arXiv:hep-ph/0701211].

\bibitem{Kiriyama:2006ui}
  O.~Kiriyama, D.~H.~Rischke and I.~A.~Shovkovy,
  %``Gluonic phase versus LOFF phase in two-flavor quark matter,''
  Phys.\ Lett.\  B {\bf 643}, 331 (2006)
  [arXiv:hep-ph/0606030];
%\bibitem{Hashimoto:2007ut}
  M.~Hashimoto and V.~A.~Miransky,
  %``Gluonic phases and phase diagram in neutral two flavor dense QCD,''
  Prog.\ Theor.\ Phys.\  {\bf 118}, 303 (2007)
  [arXiv:0705.2399 [hep-ph]].

\bibitem{Polyakov}
%\bibitem{Fukushima:2003fw}
  K.~Fukushima,
  %``Chiral effective model with the Polyakov loop,''
  Phys.\ Lett.\  B {\bf 591}, 277 (2004)
  [arXiv:hep-ph/0310121].

\bibitem{PlApp1}
%\bibitem{Ratti:2005jh}
  C.~Ratti, M.~A.~Thaler and W.~Weise,
  %``Phases of QCD: Lattice thermodynamics and a field theoretical model,''
  Phys.\ Rev.\  D {\bf 73}, 014019 (2006)
  [arXiv:hep-ph/0506234];
%\bibitem{Ratti:2007jf}
  C.~Ratti, S.~Roessner and W.~Weise,
  %``Quark number susceptibilities: Lattice QCD versus PNJL model,''
  Phys.\ Lett.\  B {\bf 649}, 57 (2007)
  [arXiv:hep-ph/0701091].

\bibitem{Megias:2004hj}
  E.~Megias, E.~Ruiz Arriola and L.~L.~Salcedo,
  %``Polyakov loop in chiral quark models at finite temperature,''
  Phys.\ Rev.\  D {\bf 74}, 065005 (2006)
  [arXiv:hep-ph/0412308].

\bibitem{PlApp2}
%\bibitem{Ghosh:2006qh}
  S.~K.~Ghosh, T.~K.~Mukherjee, M.~G.~Mustafa and R.~Ray,
  %``Susceptibilities and speed of sound from PNJL model,''
  Phys.\ Rev.\  D {\bf 73}, 114007 (2006)
  [arXiv:hep-ph/0603050];
%\bibitem{Ghosh:2007wy}
%  S.~K.~Ghosh, T.~K.~Mukherjee, M.~G.~Mustafa and R.~Ray,
  %``PNJL model with a Van der Monde term,''
  Phys.\ Rev.\  D {\bf 77}, 094024 (2008)
  [arXiv:0710.2790 [hep-ph]].

\bibitem{Hansen:2006ee}
  H.~Hansen, W.~M.~Alberico, A.~Beraudo, A.~Molinari, M.~Nardi and C.~Ratti,
  %``Mesonic correlation functions at finite temperature and density in the
  %Nambu-Jona-Lasinio model with a Polyakov loop,''
  Phys.\ Rev.\  D {\bf 75}, 065004 (2007)
  [arXiv:hep-ph/0609116].

\bibitem{PlApp4}
%\bibitem{Sasaki:2006ww}
  C.~Sasaki, B.~Friman and K.~Redlich,
  %``Susceptibilities and the phase structure of a chiral model with  Polyakov
  %loops,''
  Phys.\ Rev.\  D {\bf 75}, 074013 (2007)
  [arXiv:hep-ph/0611147].

\bibitem{PlApp3}
  W.~j.~Fu, Z.~Zhang and Y.~x.~Liu,
  %``2+1 Flavor Polyakov--Nambu--Jona-Lasinio Model at Finite Temperature and
  %Nonzero Chemical Potential,''
  Phys.\ Rev.\  D {\bf 77}, 014006 (2008)
  [arXiv:0711.0154 [hep-ph]];
%\bibitem{Ciminale:2007sr}
  M.~Ciminale, R.~Gatto, N.~D.~Ippolito, G.~Nardulli and M.~Ruggieri,
  %``Three flavor Nambu-Jona Lasinio model with Polyakov loop and competition
  %with nuclear matter,''
  Phys.\ Rev.\  D {\bf 77}, 054023 (2008)
  [arXiv:0711.3397 [hep-ph]].

\bibitem{PlApp5}
%\bibitem{Fukushima:2008wg}
  K.~Fukushima,
  %``Phase diagrams in the three-flavor Nambu--Jona-Lasinio model with the
  %Polyakov loop,''
  Phys.\ Rev.\  D {\bf 77}, 114028 (2008)
  [arXiv:0803.3318 [hep-ph]].

\bibitem{PlApp6}
%\bibitem{Sakai:2008py}
  Y.~Sakai, K.~Kashiwa, H.~Kouno and M.~Yahiro,
  %``Polyakov loop extended NJL model with imaginary chemical potential,''
  Phys.\ Rev.\  D {\bf 77}, 051901 (2008)
  [arXiv:0801.0034 [hep-ph]];
%\bibitem{Sakai:2008um}
%  Y.~Sakai, K.~Kashiwa, H.~Kouno and M.~Yahiro,
  %``Phase diagram in the imaginary chemical potential region and extended Z3
  %symmetry,''
%  Phys.\ Rev.\  D {\bf 78}, 036001 (2008)
  {\it ibid.}  D {\bf 78}, 036001 (2008)
  [arXiv:0803.1902 [hep-ph]].

\end{thebibliography}
\end{document}